\documentclass{jfm}
\usepackage{amsmath}
\usepackage{mathtools,tikz}
\usepackage{graphicx}
\usepackage{epstopdf, epsfig}
\usepackage{scalerel,stackengine}
\usepackage{subfig}
\usepackage{float}
\usepackage{bm}
\usepackage{tikz}
\usetikzlibrary{positioning}

\stackMath
\newcommand\reallywidehat[1]{%
\savestack{\tmpbox}{\stretchto{%
  \scaleto{%
    \scalerel*[\widthof{\ensuremath{#1}}]{\kern-.6pt\bigwedge\kern-.6pt}%
    {\rule[-\textheight/2]{1ex}{\textheight}}
  }{\textheight}%
}{0.5ex}}%
\stackon[1pt]{#1}{\tmpbox}%
}

\newcommand{\mat}[1]{{\mathsfbi{#1}}}
\newcommand{\mA}{\mat{A}}
\newcommand{\mB}{\mat{B}}
\newcommand{\mG}{\mat{G}}
\newcommand{\mH}{\mat{H}}
\newcommand{\Htilde}{\widetilde{\mat{H}}}
\newcommand{\mI}{\mat{I}}
\newcommand{\mL}{\mat{L}}
\newcommand{\mM}{\mat{M}}
\newcommand{\mP}{\mat{P}}
\newcommand{\mR}{\mat{R}}

\newcommand{\vect}[1]{{\bm{#1}}}
\newcommand{\vf}{\vect{f}}
\newcommand{\vg}{\vect{g}}
\newcommand{\vh}{\vect{h}}
\newcommand{\vq}{\vect{q}}
\newcommand{\vQ}{\vect{Q}}
\newcommand{\vu}{\vect{u}}

\newcommand{\vw}{\vect{w}}
\newcommand{\vx}{\vect{x}}
\newcommand{\vz}{\vect{z}}
\newcommand{\vphi}{\vect{\phi}}
\newcommand{\vpsi}{\vect{\psi}}
\newcommand{\vxi}{\vect{\xi}}

\newcommand{\ddt}{\frac{\mathrm{d}}{\mathrm{d}t}}

\begin{document}

\shorttitle{Harmonic Resolvent Analysis}
\shortauthor{A. Padovan, S. E. Otto, C. W. Rowley}

\title{Analysis of amplification mechanisms and cross-frequency interactions in nonlinear flows via the harmonic resolvent}

\author{Alberto Padovan\aff{1}
  \corresp{\email{apadovan@princeton.edu}},
  Samuel E. Otto\aff{1},
 \and Clarence W. Rowley\aff{1}}

\affiliation{\aff{1}Department of Mechanical and Aerospace Engineering, Princeton University,
Princeton, NJ 08540, USA}

\maketitle

\begin{abstract}

We propose a framework that elucidates the input-output characteristics of flows
with complex dynamics arising from nonlinear interactions between different time
scales.
More specifically, we consider a periodically time-varying base flow, and
perform a frequency-domain analysis of periodic perturbations about this
base flow; the response of these perturbations is governed by the harmonic resolvent,
which is a linear operator similar to the harmonic transfer function introduced by \citet{Wereley91}.
This approach makes it possible to explicitly capture the
triadic interactions that are responsible for the energy transfer between
different time scales in the flow.
For instance, perturbations at frequency $\alpha$ are coupled with
perturbations at frequency $\omega$ through the base flow at frequency
$\omega-\alpha$.
We draw a connection with resolvent analsyis, which is a special case of
the harmonic resolvent when evaluated about a steady base flow. We show that
the left and right singular vectors of the harmonic resolvent are the optimal response and
forcing modes, which can be understood as full spatio-temporal signals that
reveal space-time amplification characteristics of the flow. We illustrate the
method on examples, including a three-dimensional system of ordinary differential equations and the flow over an airfoil at near-stall angle of attack.  
\end{abstract}


\section{Introduction}

Model-based approaches rooted in linear systems theory have helped shed light on
the nature of energy amplification mechanisms in flows of interest. It has been
shown through linear analyses, for instance, that the transient energy growth in
channel flows is due to the non-normality of the linearized Navier-Stokes
operator governing the dynamics of perturbations about the well-known laminar
parabolic velocity profile \citep{schmid}.
\citet{MRJ2005} and \citet{BJM2010}, on the other hand, have
investigated energy amplification mechanisms in channel flows and turbulent
pipe flow by studying the linearized response to perturbations,
via {\em input-output analysis} or {\em resolvent
  analysis}.
Likewise, \citet{dawson} have
recently investigated the resonance and pseudoresonance mechanisms in low
Reynolds number cylinder flow and turbulent pipe flow using similar techniques.
Although these analyses provide valuable insight into the
amplification mechanisms of given flows, they do so under the assumption of
small amplitude fluctuations about a steady base flow (often the temporal
mean). Furthermore, because of the time-invariant nature of the chosen base
flow, such methods are inherently incapable of capturing the cross-frequency
interactions that are responsible for the energy transfer between motions at
different time scales. Not only are these cross-frequency interactions the
fundamental mechanisms behind the energy cascade in turbulent flows, but they
are also responsible for the rise of complex dynamics in laminar flows such as
boundary layers \citep{mittal2013} and mixing layers \citep{Ho1981}. The
limitations of these analyses are of course known and have partly been addressed
in the past. For instance, \citet{MRJ2008} introduced a perturbation analysis
framework to study the amplification mechanisms of linear, small-amplitude,
time-periodic systems and applied it to two-dimensional oscillating channel
flow. In a similar spirit, we develop a mathematical framework that
attempts to address the limitations of linear time-invariant analyses, while
still providing insight into the input-output characteristics of the fluid flow at
hand.

This paper considers a framework we call {\em harmonic resolvent analysis}, in which the dynamics are expanded about a
{\em periodically time-varying} base flow, which can be viewed as the large-scale,
coherent structures of the flow.  It will be shown in section \ref{sec:
  mathFormulation} that this formulation justifies treating the higher-order
terms in the expansion as small input disturbances.  Analyzing the linearized
governing equations in the frequency domain enables the explicit computation of
the harmonic resolvent, a linear operator which governs the dynamics of small
perturbations about the time-varying, periodic base flow.  Because of the
multimodal nature of the base flow, the harmonic resolvent can capture the leading-order
cross-frequency interactions, which arise in the form of triadic couplings
between perturbations at frequencies $\omega$ and $\alpha$ through the base flow
at frequency $\omega - \alpha$.  The number of triads that can be captured is
determined by the number of Fourier modes that are retained in the base flow and
we show that if the latter is simply a steady flow, then resolvent analysis is
recovered. The harmonic resolvent operator can be viewed as a special case of the harmonic transfer function introduced by \citet{Wereley91}, which maps inputs to outputs in the space of exponentially modulated periodic signals. Furthermore, it is worth mentioning that the temporal expansion of the dynamics
about the large-scale  coherent structures of the flow that we perform is similar to the
wavenumber-space expansions applied in the generalized quasi-linear
approximation introduced by \citet{marston2016} to describe the interaction
between the large and small scale of flows in the context of direct statistical
simulations.

Similarly to the spectral analysis of the resolvent operator, the singular value
decomposition of the harmonic resolvent provides insight into the amplification mechanisms of
disturbances about the time-varying base flow.  Specifically, the right and left
singular vectors of the operator are the optimal spatio-temporal forcing and the
most amplified spatio-temporal reponse patterns, respectively.  It will be shown
that one can also seek the optimal spatial forcing and most amplified spatial
response at selected frequency pairs in order to study their cross-frequency
amplification mechanisms.

In section \ref{sec: toyModel} we illustrate the method on a system of three
ordinary differential equations, whose low dimensionality and time-periodic
dynamics allow us to illustrate the characteristics of the harmonic resolvent
and to draw a natural comparison between the harmonic resolvent framework and
the usual resolvent analysis.

Finally, in section \ref{sec: airfoilSec} we consider the flow over an airfoil
at near-stall angle of attack.  This flow exhibits multi-chromatic time-periodic
dynamics, which we study using the harmonic resolvent. In particular, we
compute the optimal forcing and response modes via the singular value
decomposition of the harmonic resolvent, and we analyze the amplification mechanisms of
perturbations about the periodically time-varying base flow that arise from the
nonlinear dynamics.

\section{Mathematical formulation}
\label{sec: mathFormulation}

In this section, we define the harmonic resolvent operator, first for a general nonlinear system and then for incompressible fluid flows.

\subsection{General nonlinear system}
\label{sec:gener-nonl-syst}
We consider the nonlinear autonomous system
\begin{equation}
\label{eqn: generalSys}
    \frac{\mathrm{d}}{\mathrm{d}t}\vq(t) = \vf\big(\vq(t)\big)
\end{equation}
with state $\vq(t)$.
In three-dimensional incompressible fluid flows, the state is the
three-dimensional vector velocity field along with the scalar pressure.

In the harmonic resolvent framework, we are interested in studying the
amplification mechanisms of small perturbations $\vq'(t)$ about a time-varying base flow that is periodic with
period~$T$, given by
\begin{equation}
\label{eqn: baseFlow}
    \vQ(t) = \sum_{\omega \in \Omega_b} \hat\vQ_{\omega}e^{i\omega t}
\end{equation}
with $\Omega_b\subset \frac{2\pi}{T}\mathbb{Z}$. The base flow does not need to satisfy the governing equations and $\Omega_b$ usually contains a small subset of frequencies that approximate the dynamics of the large coherent structures present in the flow. We proceed by seeking perturbations of the form
\begin{equation}
    \vq'(t) = \sum_{\omega \in \Omega}\hat\vq'_{\omega}e^{i\omega t}
\end{equation}
with $\Omega \subset \frac{2\pi}{T}\mathbb{Z}$. Usually $\Omega_b \subset \Omega$, as $\Omega$ is the set of temporal frequencies associated with the flow structures that one wishes to resolve. Upon substituting the decomposition $\vq(t)=\vQ(t)+\vq'(t)$ in (\ref{eqn: generalSys}) we obtain
\begin{equation}
\label{eqn: decomposedSys}
    \frac{\mathrm{d}}{\mathrm{d}t}\vq'(t) = \underbrace{\mathcal{D}_{\vq}\vf\left(\vQ(t)\right)}_{\mA(t)}\vq'(t) + \vh'(t)
\end{equation}
where 
\begin{equation}
  \label{eq:1}
    \vh'(t) = \left[-\frac{\mathrm{d}}{\mathrm{d}t}\vQ(t) + \vf\big(\vQ(t)\big)\right] + o\left(\|\vq'(t)\|\right).
\end{equation}
The first term in~\eqref{eq:1} is the error associated with the base flow not satisfying the
dynamics, while the second represents the higher order terms in the dynamics. Before proceeding further, we observe
that $\mA(t)$ is periodic with period~$T$
(since $\vQ$ is periodic), and hence it can be
represented in terms of a Fourier series, analogous to~\eqref{eqn:
  baseFlow}. We then obtain the following expression for the perturbation at frequency~$\omega$:
\begin{equation}
  \label{eqn: freqDomain_Equation}
  i\omega \hat\vq'_{\omega} = \sum_{\alpha\in \Omega}
  \hat\mA_{\omega-\alpha} \hat\vq'_{\alpha}
  + \hat\vh'_{\omega} \qquad \forall \omega \in \Omega.
\end{equation}
We neglect frequencies $\omega$ that are not in $\Omega$.
For ease of notation, let $\hat\vq'$ be the vector of
$\hat\vq'_\omega$ for all frequencies $\omega \in \Omega$, and let
$\hat\vh'$ be defined similarly.  We define
the harmonic resolvent $\mH$ by
\begin{equation}
  \label{eq:2}
  \big[\mH^{-1}\hat \vq'\Big]_\omega = i\omega \hat \vq_\omega' - \sum_{\alpha\in\Omega}
  \hat \mA_{\omega-\alpha}\hat \vq_\alpha',
\end{equation}
and formula~\eqref{eqn: freqDomain_Equation} may be written as
\begin{equation}
  \label{eq:4}
  \hat\vq' = \mH\hat\vh'.
\end{equation}
The harmonic resolvent $\mH$ is thus a linear operator that
describes the dynamics of small periodic perturbations $\hat\vq'$ about a
periodic base flow, in response to a periodic input forcing $\hat\vh'$.
Note that, if external inputs are present (such as a control input, or noise forcing the system),
these enter into the formulation in the same way that $\vh'$ does, and this leads to the harmonic transfer function of~\citet{Wereley91}.

A more thorough discussion of the characteristics of the harmonic resolvent
is given at the end of section~\ref{sec:incompressible-fluids}.

\subsection{Bilinear system: incompressible fluid flow}
\label{sec:incompressible-fluids}
We now consider an incompressible fluid flow governed by the Navier-Stokes equations, given by
\begin{align}
\label{eqn: NS}
\begin{split}
    \frac{\partial}{\partial t}\vu + \vu\cdot \nabla \vu &= -\nabla p + \Rey^{-1}\nabla^2 \vu \\
    \nabla \cdot \vu &= 0.
\end{split}
\end{align}
Here, $\vu(\vx,t)$ and $p(\vx,t)$ are the velocity and pressure, respectively,
over the spatial domain $\mathcal{X}\subseteq \mathbb{R}^3$. For ease of notation, we will
drop the explicit dependence on~$\vx$ from here on. Equation~(\ref{eqn: NS}) can be written compactly as
\begin{equation}
\label{eqn: fluidSys}
    \frac{\partial}{\partial t}
    \underbrace{\begin{bmatrix}
    \mI & 0 \\ 0 & 0
    \end{bmatrix}}_{\mM} \begin{bmatrix}\vu(t)\\ p(t)\end{bmatrix} = \underbrace{\begin{bmatrix}
    \Rey^{-1}\nabla^2 & -\nabla \\ -\nabla \cdot & 0
    \end{bmatrix}}_{\mL} \begin{bmatrix}\vu(t)\\ p(t)\end{bmatrix} + \underbrace{\begin{bmatrix}
    -\vu(t)\cdot \nabla \vu(t) \\ 0
    \end{bmatrix}}_{\vg\left(\vu(t),\vu(t)\right)}.
\end{equation}
We denote the state vector by $\vq=\left(\vu,p\right)$, and consider
perturbations about a periodic base flow, as in Section~\ref{sec:gener-nonl-syst}:
\begin{equation}
    \vq(t) = \vQ(t)+\vq'(t) = \sum_{\omega \in \Omega_b}\hat\vQ_{\omega}e^{i\omega t} + \sum_{\omega \in \Omega}\hat\vq'_{\omega}e^{i\omega t} \label{eqn: Qflow},
\end{equation}
where $\Omega_b \subseteq \Omega \subset \frac{2\pi}{T}\mathbb{Z}$.  As before,
we seek an input-output representation for the perturbations $\vq'$. Substituting (\ref{eqn: Qflow}) in (\ref{eqn: fluidSys}), and neglecting frequencies $\omega \notin \Omega$, we obtain
\begin{equation}
\label{eqn: fluidSys_omegas}
    i\omega \mM \hat\vq'_{\omega} = \mL \hat\vq'_{\omega} + \sum_{\alpha \in \Omega}\left(\vg(\hat\vQ_{\omega-\alpha},\hat\vq'_{\alpha}) +  \vg(\hat\vq'_{\alpha},\hat\vQ_{\omega-\alpha})\right) + \hat\vh'_{\omega},\qquad \forall \omega \in \Omega
\end{equation}
where, as in the previous section, $\hat\vh'_{\omega}$ is the Fourier
mode of the base flow error along with the terms that are nonlinear in
$\hat\vq'_\omega$ (see formula~\eqref{eq:1}).
We again let $\hat\vq'$ denote the vector of $\hat\vq_\omega'$ for all
frequencies $\omega\in\Omega$, and define the harmonic resolvent
$\mH$ by
\begin{equation}
  \label{eq:5}
  \big[\mH^{-1}\hat\vq'\big]_\omega =
  (i\omega \mM - \mL)\hat\vq'_\omega -\sum_{\alpha\in\Omega}\big[\vg(\hat\vQ_{\omega-\alpha},\hat\vq'_{\alpha}) +  \vg(\hat\vq'_{\alpha},\hat\vQ_{\omega-\alpha}) \big].
\end{equation}
Finally, formula~\eqref{eqn: fluidSys_omegas} may be written compactly as
\begin{equation}
    \hat\vq' = \mH\hat\vh'.
\end{equation}
As specified at the end of section \ref{sec:gener-nonl-syst}, inputs such as a control signal or an external disturbance enter in the system in the same way as $\vh'$.

At this point, a few comments on the structure of the harmonic resolvent operator
are in order.  First, note that the number of frequencies
in the set $\Omega$ may be infinite (e.g., $\Omega=\frac{2\pi}{T}\mathbb{Z}$),
in which case the harmonic resolvent is an infinite dimensional
operator.  However, in practice, one truncates its dimensionality by selecting a
finite number of frequencies $\omega \in \Omega$ that are considered to be of
interest. The dimension of $\mH$ is therefore proportional to the number of
frequencies in~$\Omega$.  As mentioned previously, we usually consider the
frequencies $\Omega_b$ in the base flow to be a subset of the frequencies
$\Omega$ of the perturbations. That is, one usually wishes to study the dynamics
of perturbations at multiple frequencies, about a filtered representation of the
large scale structures that are observed in the flow. The number of frequencies
$\omega \in \Omega_b$ affects the accuracy of the linear operator in
representing the nonlinear dynamics of the flow. This becomes clear once we
observe from formula~\eqref{eqn: fluidSys_omegas} that  perturbations at different temporal
frequencies are linearly coupled to one another via the base flow. More
precisely, structures at frequency $\omega$ are coupled to structures at
frequency $\alpha$ through the base flow at the frequency difference $\omega -
\alpha$. Of course, $\omega-\alpha$ needs to be in $\Omega_b$ if one wishes to
capture the aforementioned interaction. For instance, if the base flow has
frequencies $\Omega_b = \{-\omega,0,\omega\}$ and we want to study the dynamics of
perturbations over the set of frequencies $\Omega = \{-2\omega,-\omega,0,\omega,2\omega\}$, then
the inverse of the harmonic resolvent takes the form
\begin{equation*}
    \mH^{-1} = \begin{bmatrix}
    \mR^{-1}_{-2\omega} & \mG_{-\omega} & 0 & 0 & 0\\
    \mG_\omega & \mR^{-1}_{-\omega} & \mG_{-\omega} & 0 & 0\\
    0 & \mG_\omega & \mR^{-1}_{0} & \mG_{-\omega} & 0\\
    0 & 0 & \mG_\omega & \mR^{-1}_\omega & \mG_{-\omega}\\
    0 & 0 & 0 & \mG_\omega & \mR^{-1}_{2\omega}\\
    \end{bmatrix}
\end{equation*}
where
\begin{align*}
    \mG_\omega \vq &= -\vg(\hat\vQ_\omega,\vq) - \vg(\vq,\hat\vQ_\omega)\\
    \mR^{-1}_\omega &= i\omega \mM - \mL - \mG_0.
\end{align*}
Note that $\mR_\omega$ is the usual resolvent operator at frequency~$\omega$ (i.e., the resolvent of the operator linearized about the constant base flow $\hat\vQ_0$).
In fact, in the special case that the base flow is constant (i.e., $\Omega_b=\{0\}$), the harmonic resolvent becomes block diagonal, and perturbations at different frequencies are decoupled.

\subsection{Global amplification mechanisms from the harmonic resolvent}
\label{subsec: globalAmp}
The mathematical formulation presented in the previous sections leads to a
linear time-periodic input-output system in Fourier space, represented by
\begin{equation}
    \hat\vq' = \mH\hat\vw'
\end{equation}
where the harmonic resolvent~$\mH$ governs the dynamics of
perturbations about a periodic base flow $\hat\vQ$ in response to some periodic
forcing $\hat\vw'$. There are several ways to view the perturbation $\hat\vw'$. From the
point of view of control theory, $\hat\vw'$ can be interpreted as an external
input, which might be chosen to achieve some control objective. In a more
physics-driven approach, $\hat\vw'$ can be understood as the frequency-domain
representation of the nonlinearities that feed back into the linear harmonic
resolvent. Alternatively, $\hat\vw'$ can be viewed as an external disturbance that perturbs the system around a known periodic orbit. 

In any of these circumstances, one may want to understand the dominant mechanisms by
which space-time inputs are amplified through $\mH$. One way to do so is by
seeking a unit-norm space-time input $\hat\vw'$ that leads to the most amplified
space-time response $\hat\vq'$.

Before proceeding to find such an ``optimal'' forcing, we mention a subtlety: in
particular, we are not interested in
perturbations that only serve to shift the phase of the original periodic
orbit.  For instance, if the perturbation is $\vq'(t)=\ddt \vQ(t)$, where
$\vQ(t)$ is the periodic base flow, then for small~$\epsilon$, we have
\begin{equation*}
  \vQ(t) + \varepsilon\vq'(t) = \vQ(t+\varepsilon) + O(\varepsilon^2),
\end{equation*}
so such a perturbation merely shifts the phase of the periodic orbit.  We are
therefore interested only in perturbations $\vq'$ that are orthogonal to
$\ddt \vQ$.  This orthogonality also holds in the frequency domain (thanks to
Parseval's identity), so in the frequency domain, we restrict ourselves to
perturbations $\hat\vq'$ orthogonal to
$\reallywidehat{\frac{\mathrm{d}}{\mathrm{d}t} \vQ} =
\{i\omega\hat\vQ_{\omega}\}_{\omega \in \Omega_b}$.

With this in mind, in order to find the perturbations that cause the greatest
amplification, we wish to solve the following optimization problem:
\begin{align}
\label{eqn: optimizationProb}
    \begin{split}
        \max_{\hat\vw'}\quad &\langle\mH\hat\vw', \mH\hat\vw'\rangle \\
        \text{subject to: } &\langle \hat\vw',\hat\vw'\rangle = 1 \\
        & \langle \mH\hat\vw',\reallywidehat{\frac{\mathrm{d}}{\mathrm{d}t} \vQ} \rangle = 0
    \end{split}
\end{align}
In order to approach the solution of the optimization problem we define a
unit-norm vector $\vz$ in the direction of
$\mH^*\big[\reallywidehat{\frac{\mathrm{d}}{\mathrm{d}t} \vQ} \big]$ (where
$\mH^*$ denotes the adjoint of $\mH$), and an orthogonal projection $\mP
= \mI - \vz\vz^*$ that serves the purpose of projecting out the component of the
forcing that would merely phase-shift the base flow. The
problem~\eqref{eqn: optimizationProb} is therefore equivalent to
\begin{align}
\label{eqn: optimizationProb2}
    \begin{split}
        \max_{\hat\vw'}\quad &\langle\mH\mP\hat\vw', \mH\mP\hat\vw'\rangle \\
        \text{subject to: } &\langle \hat\vw',\hat\vw'\rangle = 1 \\
    \end{split}
\end{align}
In order to reflect our choice of neglecting the possibility of merely shifting
along the base flow, we revise our definition of harmonic transfer function to
$\Htilde = \mH\mP$, whose range is orthogonal to $\reallywidehat{\frac{\mathrm{d}}{\mathrm{d}t} \vQ}$. Finally, it can be shown that (\ref{eqn: optimizationProb2}) leads to the eigenvalue problem 

\begin{equation}
\label{eqn: globalEvalProblem}
    \Htilde^{*}\Htilde\hat\vpsi = \sigma \hat\vpsi
\end{equation}
where the optimal unit-norm forcing $\hat\vpsi$ is the first right singular
vector of $\Htilde$, and $\sigma$ is the largest singular value. If we left-multiply (\ref{eqn: globalEvalProblem}) by $\Htilde$ and define $\hat\vphi = \Htilde\hat\vpsi$ we obtain
\begin{equation}
    \Htilde\Htilde^*\hat\vphi = \sigma \hat\vphi
\end{equation}
from which we can conclude that the optimal (most amplified) response,
$\hat\vphi$, is the corresponding left singular vector of $\Htilde$. We will
refer to the right singular vectors of~$\Htilde$ as {\em input modes} and we will
refer to the left singular vectors as {\em output modes}.

Proceeding further, the response of the linear time-periodic system to an
arbitrary input $\hat\vw'$ can be expressed as a linear combination of input and
output modes of the harmonic resolvent, as follows:
\begin{equation}
\label{eqn: inputOutput}
    \hat\vq' = \Htilde\hat\vw'=\sum_{j = 1}^{N-1}\sigma_j\hat\vphi_j\underbrace{\hat\vpsi_j^*\hat\vw'}_{\langle\hat\vw', \hat\vpsi_j \rangle}
\end{equation}
where $\hat{\vphi}'_j, \hat{\vpsi}'_j \in \mathbb{C}^N$. 
Observe that we sum to $N-1$ as we have constrained the range of $\Htilde$ to a $(N-1)$-dimensional subspace orthogonal to $\reallywidehat{\frac{\mathrm{d}}{\mathrm{d}t} \vQ}$. Equation (\ref{eqn: inputOutput}) sheds some light on the information contained
within the output and input modes. In particular, the output modes~$\hat\vphi_j$
form an orthonormal basis for the range of~$\Htilde$ and identify the
spatio-temporal structures that are preferentially excited in
response to some external input. The input modes~$\hat\vpsi_j$ form an
orthonormal basis for the domain of~$\Htilde$ and identify the spatio-temporal
structures that are most effective at exciting an energetic response.
That is, the input modes relate to the spatio-temporal sensitivity of the
flow to external inputs. This concept can be easily understood in terms of the
inner product in the underbrace of (\ref{eqn: inputOutput}). For the sake of
example, let us consider a rank-1
approximation of $\Htilde$ and assume that $\sigma_1 \approx
  1$. If the external input $\hat\vw'$ aligns poorly with the input
mode $\hat\vpsi_1$, then $\langle\hat\vw', \hat\vpsi_1 \rangle \ll 1$ and
consequently $\|\hat\vq'\|\ll 1$, meaning that $\hat\vw'$ is not effective at
exciting a very energetic response through the harmonic resolvent. On
the other hand, if the external input aligns well with $\hat\vpsi_1$, then
$\langle\hat\vw', \hat\vpsi_1 \rangle \approx \|\hat\vw'\|$.  Consequently
$\hat\vq' \approx \hat\vphi_1 \|\hat\vw'\|$ and
$\|\hat\vq'\|\approx \|\hat\vw'\|$. In this case $\hat\vw'$ is (close to)
optimal and it excites the (close to) optimal most energetic response.
Understanding the sensitivity information contained within the input modes is
especially important if one is interested in controlling the flow. For instance, if
$\hat\vw'$ is a chosen control input, it is advisable to design it in such a way
that $\langle \hat\vw',\hat\vpsi_j\rangle$ is maximized.

The singular values~$\sigma_j$ can be understood as the gains on the input-output pairs
$\hat\vpsi_j$,$\hat\vphi_j$, and they provide information about the rank of the
harmonic resolvent. For instance, if $\sigma_j$ is very small, then
the corresponding modes have little effect on the input-output response and can
be neglected.  Often the effective rank~$r$ of~$\Htilde$ (i.e., the number of singular
values that exceed some threshold) is such that
$r \ll N-1$, and~$\Htilde$ may be approximated as
\begin{equation}
\label{eqn: inputOutput_lowRank}
    \Htilde \approx\sum_{j = 1}^{r}\sigma_j\hat\vphi_j\hat\vpsi_j^*.
\end{equation}

\subsection{Cross-frequency amplification mechanisms from the harmonic resolvent}

The global analysis that was carried out in the previous section can be easily
extended to selected frequency pairs or selected subsets of $\Omega$. Since the
harmonic resolvent accounts for the coupling between different
frequencies, we may ask which cross-frequency interactions are most
significant. More precisely, for given $\alpha,\omega \in \Omega$, we can seek the
unit-norm forcing at frequency $\alpha$ that triggers the most amplified
response at frequency $\omega$. This corresponds to the following optimization problem:
\begin{align}
    \begin{split}
        \max_{\hat\vw'_{\alpha}}\quad  &\langle \hat\vq'_{\omega},\hat\vq'_{\omega} \rangle \\
        \text{subject to: } &\langle \hat\vw'_{\alpha},\hat\vw'_{\alpha} \rangle = 1
    \end{split}
\end{align}
where 
\begin{equation}
    \hat\vq'_{\omega} = \Htilde_{\omega,\alpha}\hat\vw'_{\alpha}.
\end{equation}
It can be shown that the optimal input $\hat\vw'_{\alpha}$ and the optimal
output $\hat\vq'_{\alpha}$ are, respectively, the first right singular vector
and the first left singular vector of $\Htilde_{\omega,\alpha}$, where the
latter is the block of~$\Htilde$ that couples structures at frequency $\omega$
with structures at frequency~$\alpha$. The corresponding singular value,
$\sigma_{\omega,\alpha}$, is the gain on the $\omega,\alpha$ cross-frequency
pair. For different values of $\omega,\alpha\in\Omega$, the magnitudes of $\sigma_{\omega,\alpha}$ provide a measure to determine which
cross-frequency couplings are responsible for the development of the structures
that are observed in the full nonlinear flow.

\section{Application to a 3-dimensional toy model}
\label{sec: toyModel}
The objective of this section is to illustrate, through a simple model, the
benefits of using the harmonic resolvent framework to analyze fluid
flows that exhibit features that arise from nonlinear mechanisms. The signature
of such flows is a non-monochromatic energy spectrum, which highlights the
action of the nonlinear term in distributing energy across selected
frequencies. We consider a 3-dimensional system of ordinary differential
equations defined as follows:
\begin{align}
    \label{eqn: 3odes}
    \begin{split}
        \dot x &= {\mu} x - {\gamma} y - {\alpha} x z - {\beta} x y \\
        \dot y &= {\gamma} x + {\mu} y - {\alpha} y z + {\beta} x^2 \\
        \dot z &= -{\alpha} z + {\alpha} \left(x^2 + y^2 \right),
    \end{split}
\end{align}
where $\dot x$ denotes $\mathrm{d}x/\mathrm{d}t$, and $\alpha,\gamma,\mu>0$.
A simple rescaling of time allows us to take $\gamma=1$ without loss of
generality, so henceforth we assume $\gamma=1$.

Although we do not claim that this toy model represents any specific fluid flow,
it does share some features with the Navier-Stokes equations.  First, like the
Navier-Stokes equations, the
nonlinearities are quadratic and energy-conserving.  Recall that a dynamical
system $\ddt \vq = \vf(\vq)$ is energy conserving if
\begin{equation}
    \frac{\mathrm{d}}{\mathrm{d}t} \tfrac{1}{2}\|\vq\|^2 = \langle \vf\left(\vq \right),\vq \rangle = 0.
\end{equation}
For the Navier-Stokes equations, with typical boundary conditions on $\vu$ (e.g., $\vu=\vect{0}$ on
the boundary, or $\vu$ tangent to the boundary), one finds $\langle\vu\cdot
\nabla \vu,\vu \rangle = 0$, so the nonlinear terms in~\eqref{eqn: NS} are
energy conserving.  Similarly, for our toy model, the nonlinear terms
$\vf(x,y,z)=\big(-\alpha xz-\beta xy,-\alpha yz+\beta x^2,\alpha(x^2+y^2)\big)$
satisfy $\vf(\vq)\cdot\vq=0$, and hence are energy conserving.  In addition, we
remark that if $\beta=0$, then the system~\eqref{eqn: 3odes} is the same
as the reduced-order model of the flow past a cylinder used by
\citet{noack2003}, and is closely related to the well-known Stuart-Landau model \citet{stuart58}.

It is useful to transform the model~\eqref{eqn: 3odes} to polar coordinates:
with $x=r\cos\theta$ and $y=r\sin\theta$, the dynamics become
\begin{subequations}
  \begin{align}
  \dot r &= (\mu-\alpha z)r  \label{eq:6}\\
  \dot \theta &= 1 + \beta r\cos\theta \label{eq:7}\\
  \dot z &= \alpha(r^2-z).\label{eq:8}
  \end{align}
\end{subequations}
In these coordinates, it is clear that if $\beta^2<\alpha/\mu$, there is a limit
cycle at $r^2=z=\mu/\alpha$.  Furthermore, by integrating~\eqref{eq:7}, we
find that the period of the limit cycle is $T =2\pi/\sqrt{1-\beta^2\mu/\alpha}$,
so the fundamental frequency of the limit cycle is
\begin{equation}
  \label{eq:9}
  \omega=\sqrt{1-\beta^2\mu/\alpha}.
\end{equation}
We proceed by briefly analyzing how the dynamics of the system change as one
varies the parameter $\beta$.

\begin{figure}
\centering
\subfloat
{
\includegraphics[trim = 0.6in 0.1in 1.3in 0.1in, clip,width=0.45\textwidth]{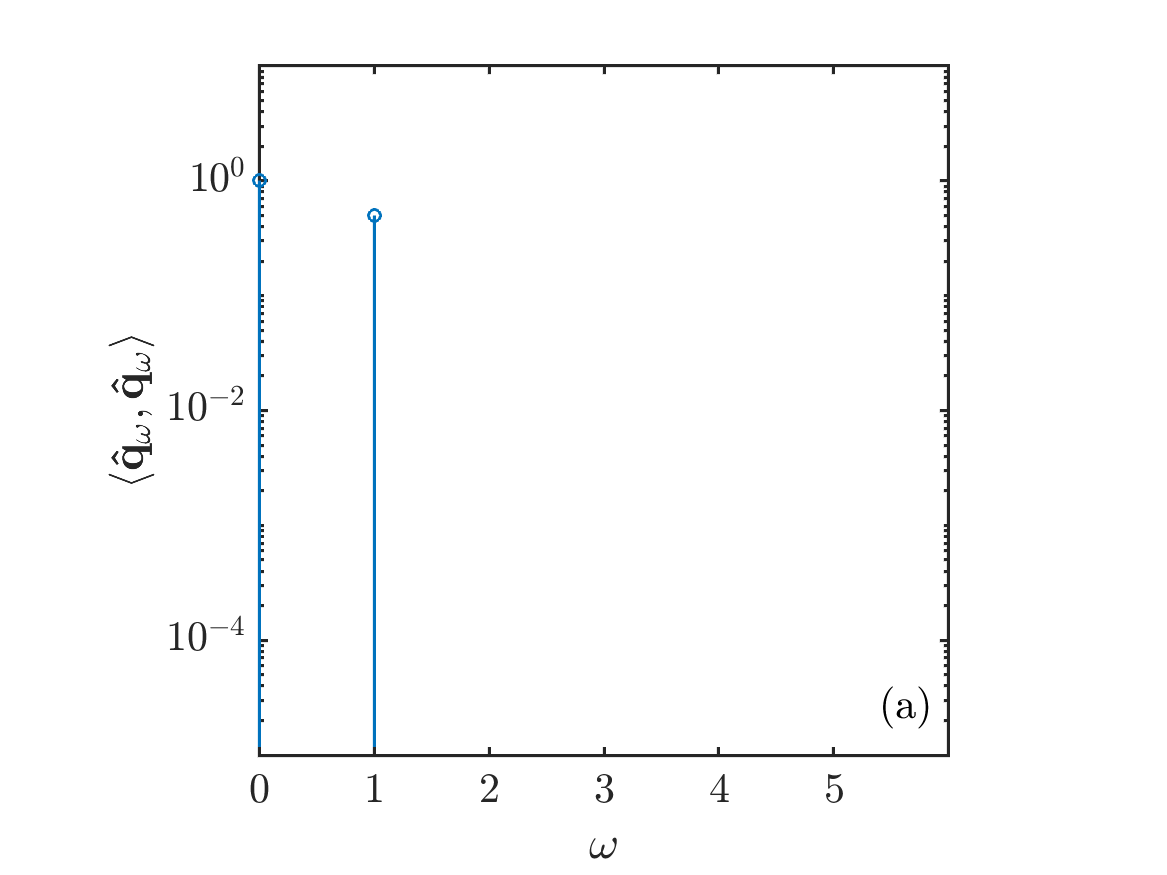}
}
\hspace{1ex}
\subfloat
{
\includegraphics[trim = 0.6in 0.1in 1.3in 0.1in, clip,width=0.45\textwidth]{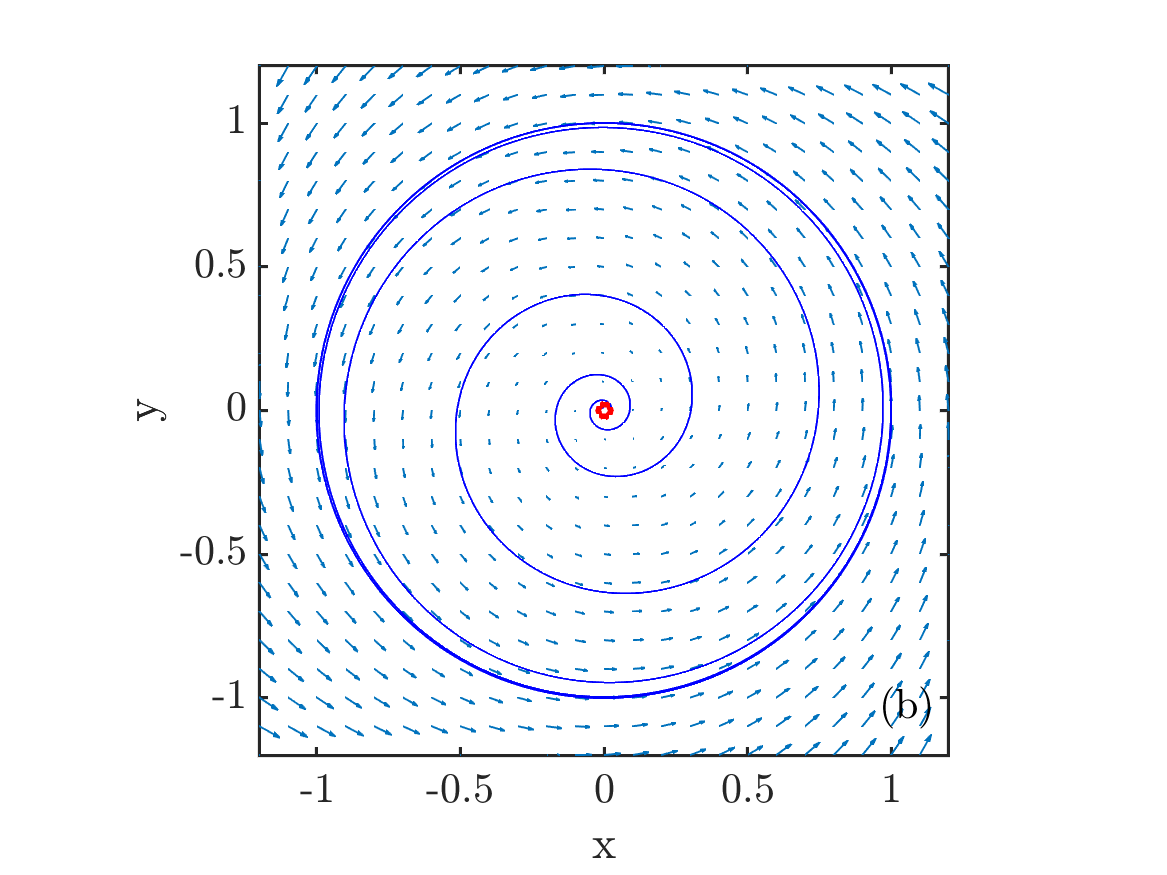}
}
\caption{Results for the toy problem~\eqref{eqn: 3odes} with $\beta=0$ and $\mu=\alpha=1/5$, showing (a) energy spectrum;
    and (b) (projected) phase portrait, with initial condition
  $\vq = (0,0.01,\mu/\alpha)$. The marker located at $(x,y) = (0,0)$ in (b)
  indicates the temporal mean of the limit cycle.}
\label{fig: beta0}
\end{figure}

When $\beta = 0$, the dynamics in the $\theta$ direction become $\dot\theta=1$,
so the system is rotationally symmetric about the $z$-axis.  Moreover, the limit
cycle is monochromatic, with frequency $\omega=1$.  Figure~\ref{fig: beta0}
shows the phase portrait and the energy spectrum of the limit cycle, for
$\mu=\alpha=1/5$ and $\beta=0$.

\begin{figure}
\centering
\subfloat
{
\includegraphics[trim = 0.6in 0.1in 1.3in 0.1in, clip,width=0.45\textwidth]{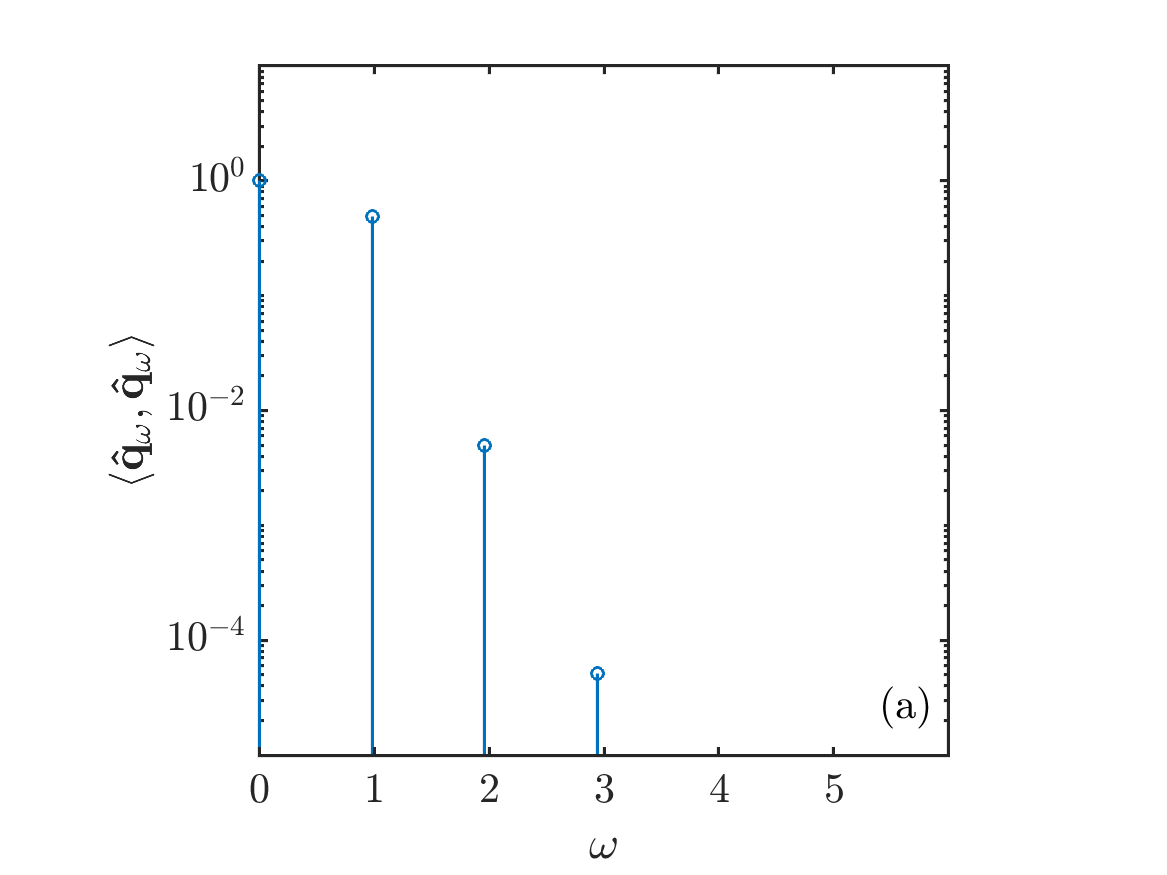}
}
\hspace{1ex}
\subfloat
{
\includegraphics[trim = 0.6in 0.1in 1.3in 0.1in, clip,width=0.45\textwidth]{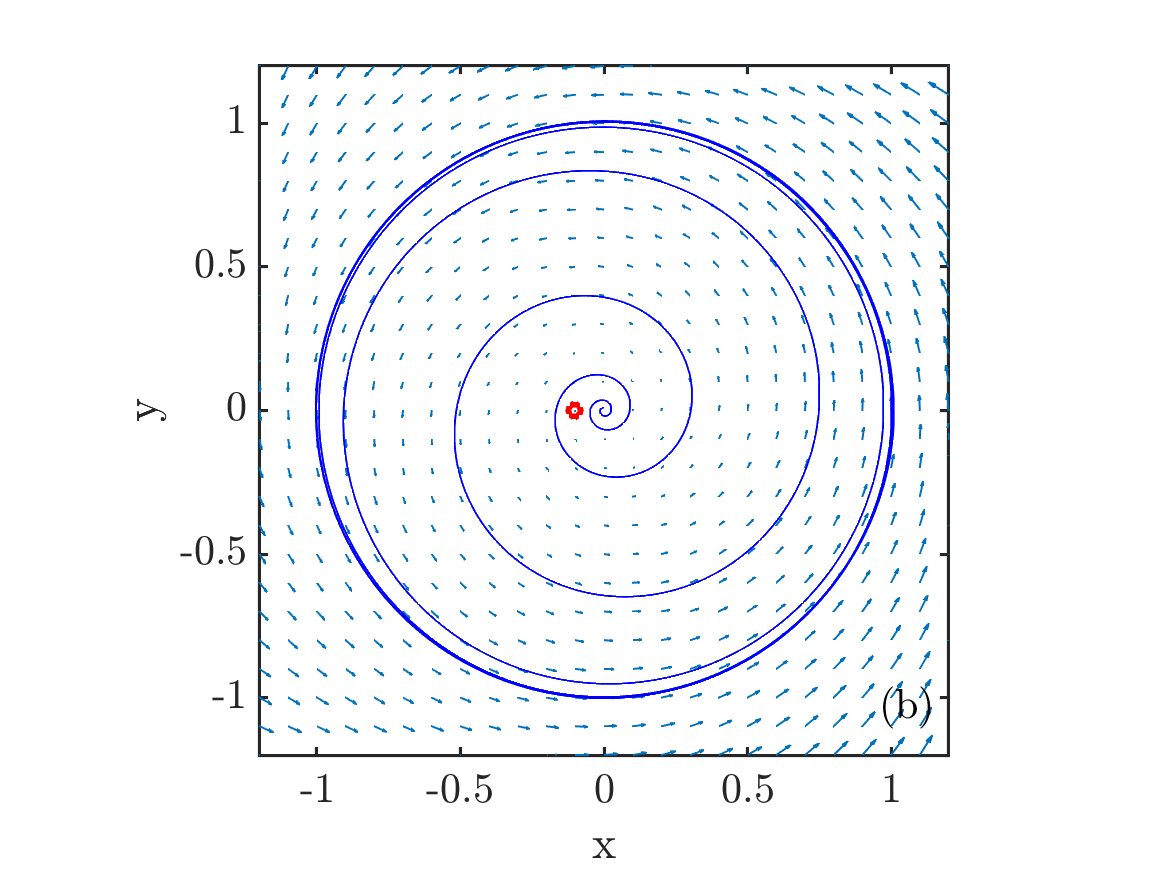}
}
\caption{Analog of Figure~\ref{fig: beta0} for $\beta=1/5$, showing (a) energy
  spectrum, and  (b) phase portrait.  Higher harmonics are present in the
  energy spectrum, and the temporal mean is shifted away from 0.}
\label{fig: beta}
\end{figure}

Next, we consider the dynamics for $0<\beta<\sqrt{\alpha/\mu}$.  There is still
a limit cycle at $r^2=z=\mu/\alpha$, but now we see from formula~\eqref{eq:7}
that there is an asymmetry: when $x>0$, the angular speed $\dot\theta$
increases, and when $x<0$, the angular speed decreases.  This will cause the
state to spend more time on the left half of the limit cycle, and so the temporal
mean is shifted to the left, as shown in Figure~\ref{fig: beta}b.  In addition,
multiple harmonics are introduced into the frequency spectrum, as shown in
Figure~\ref{fig: beta}a.  Note also that the fundamental frequency of the limit
cycle is now slightly less than~1 ($\omega\approx 0.98$), according
to~\eqref{eq:9}.

\subsection{Comparison between the harmonic resolvent framework and resolvent analysis}
In this section we compare the effectiveness of different linearizations in
predicting the response of the nonlinear system to some external periodic
forcing. For this purpose, we introduce forcing to (\ref{eqn: 3odes}) with parameters $\mu = \alpha =\beta = 1/5$ 
\begin{equation}
    \label{eqn: 3odes_force}
    \begin{aligned}
        \dot x &= \mu x -  y - \alpha x z - \beta x y +  w_1'\\
        \dot y &=  x + \mu y - \alpha y z + \beta x^2 + w_2'\\
        \dot z &= -\alpha z + \alpha \left(x^2 + y^2 \right)
    \end{aligned}{}
\end{equation}
and define $\vw'(t)=\left(\varepsilon \sin(\omega t), \varepsilon
  \cos(\omega t)\right)$ as our external periodic forcing at the
fundamental frequency $\omega = 0.98$ (from formula~(\ref{eq:9})) and with $\varepsilon = 1/5$.
The forcing $\vw'(t)$ can be expanded in a Fourier series 

\begin{equation}
    \vw'(t) = \sum_{k \in \{-1,1\}} \hat\vw'_{k}e^{i k\omega t}
\end{equation}
and the dynamics of perturbations $\vq'(t)$ about a given base flow in response to $\vw'(t)$ can be written as
\begin{equation}
    \hat\vq' = \Htilde\mB\hat\vw'
\end{equation}
where $\Htilde$ is the harmonic resolvent evaluated about the chosen base flow and $\mB$ is a block-diagonal operator through which the input $\hat\vw'$ enters the system. Throughout this section we consider perturbations with spectral energy content up to the $7^{\mathrm{th}}$ harmonic of the fundamental frequency 
\begin{equation}
  \label{eq:10}
    \vq'(t) = \sum_{k = -7}^{7}\hat\vq_{k}e^{ik\omega t}, \quad \Omega = \{-7\omega,-6\omega, \ldots ,7\omega \}
\end{equation}
and we compare the predictions obtained by linearizing about the temporal mean
$\Omega_b = \{0\}$ (see the marker in figure \ref{fig: beta}b) to predictions
obtained by linearizing about the exact limit cycle
$\Omega_b = \{-3\omega,\ldots,3\omega\}$ (see figure \ref{fig: beta}a). (Recall
that linearizing about the mean is equivalent to performing resolvent analysis.)
The results are compared against a ground truth computed by numerical
integration of~(\ref{eqn: 3odes_force}).

\begin{figure}
\centering
\subfloat
{
\includegraphics[trim = 0.5in 0in 0.6in 0.1in, clip,width=0.45\textwidth]{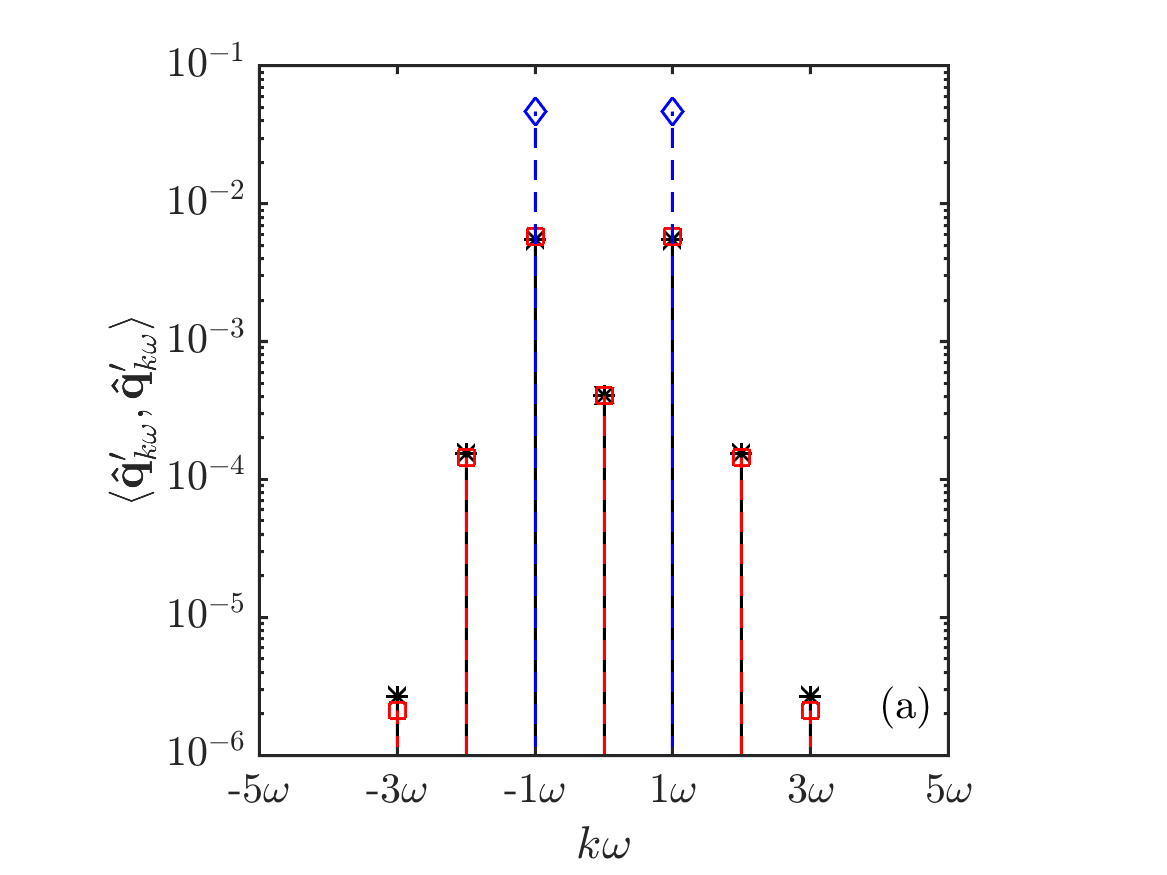}
}
\hspace{-1ex}
\subfloat
{
\includegraphics[trim = 0in 0in 0.6in 0in, clip,width=0.49\textwidth]{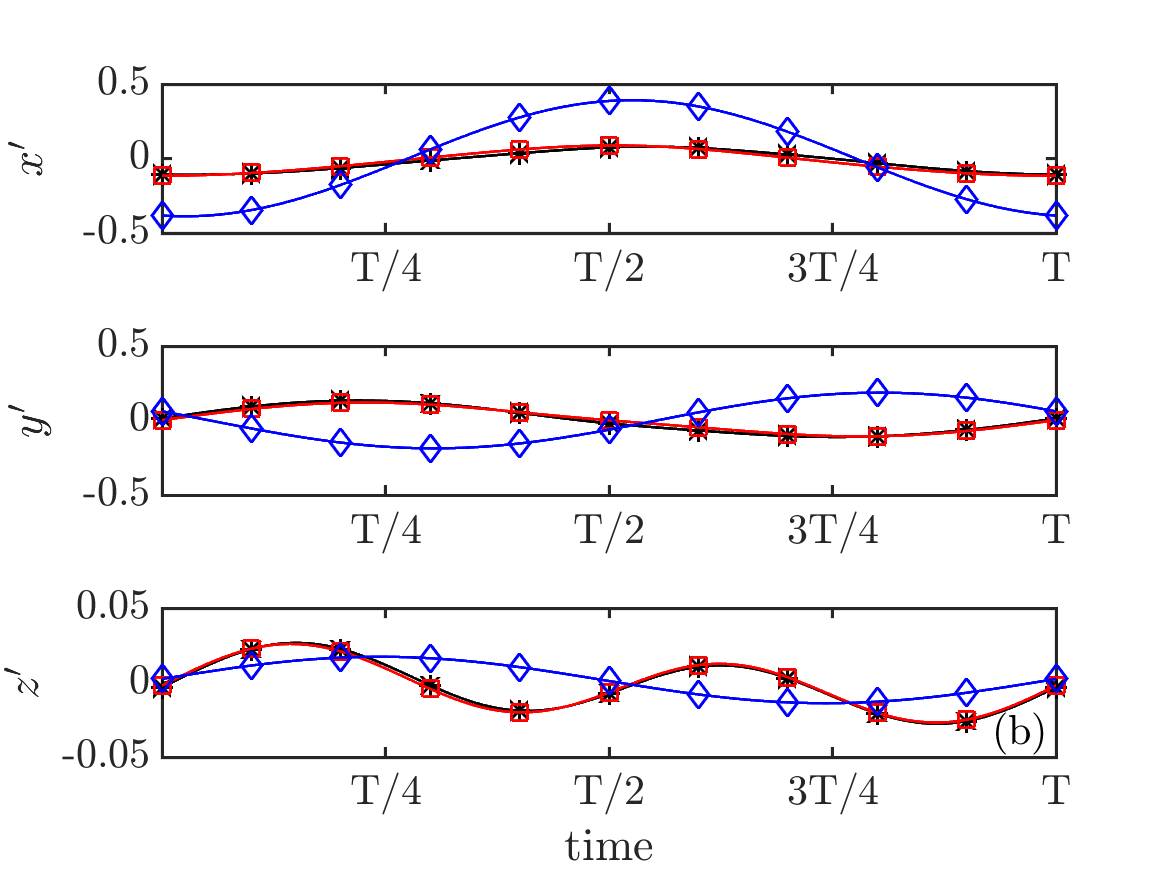}
}
\caption{Response of the system~\eqref{eqn: 3odes_force} to periodic forcing at the fundamental frequency $\omega = 0.98$. (a) spectrum of the perturbation $\mathbf{q'}(t)$, and (b) state evolution over one period $T = 2\pi/\omega$. (\protect $\ast$, \textit{Ground truth computed by numerical integration of (\ref{eqn: 3odes_force})}; $\square$, \textit{Harmonic resolvent prediction with $\Omega_b = \{-3\omega,\ldots,3\omega\}$}; $\diamond$, \textit{Harmonic resolvent prediction with $\Omega_b = \{0\}$} (equivalent to resolvent analysis)).}
\label{fig: response_toyModel}
\end{figure}

\begin{figure}
\centering
\subfloat
{
\includegraphics[trim = 0.5in 0in 0.6in 0.1in, clip,width=0.45\textwidth]{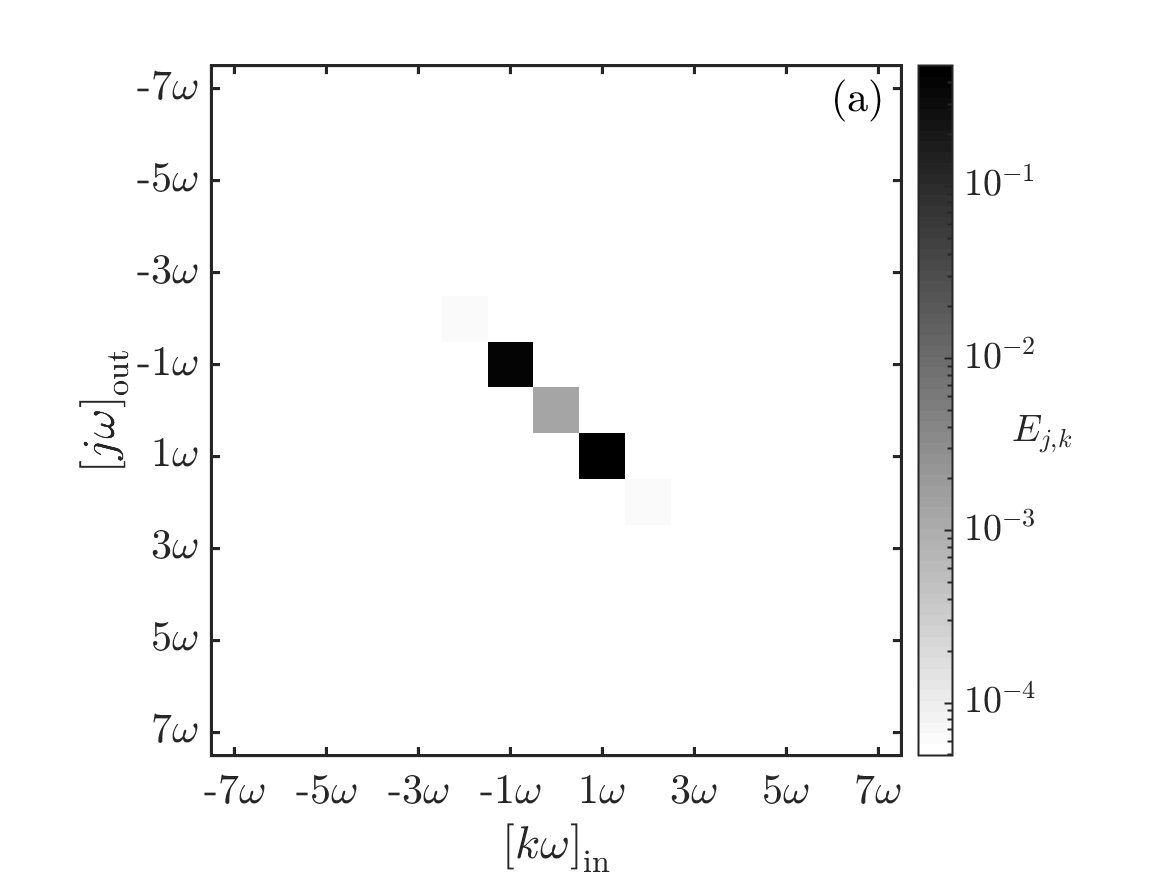}
}
\hspace{1ex}
\subfloat
{
\includegraphics[trim = 0.5in 0in 0.6in 0.1in, clip,width=0.45\textwidth]{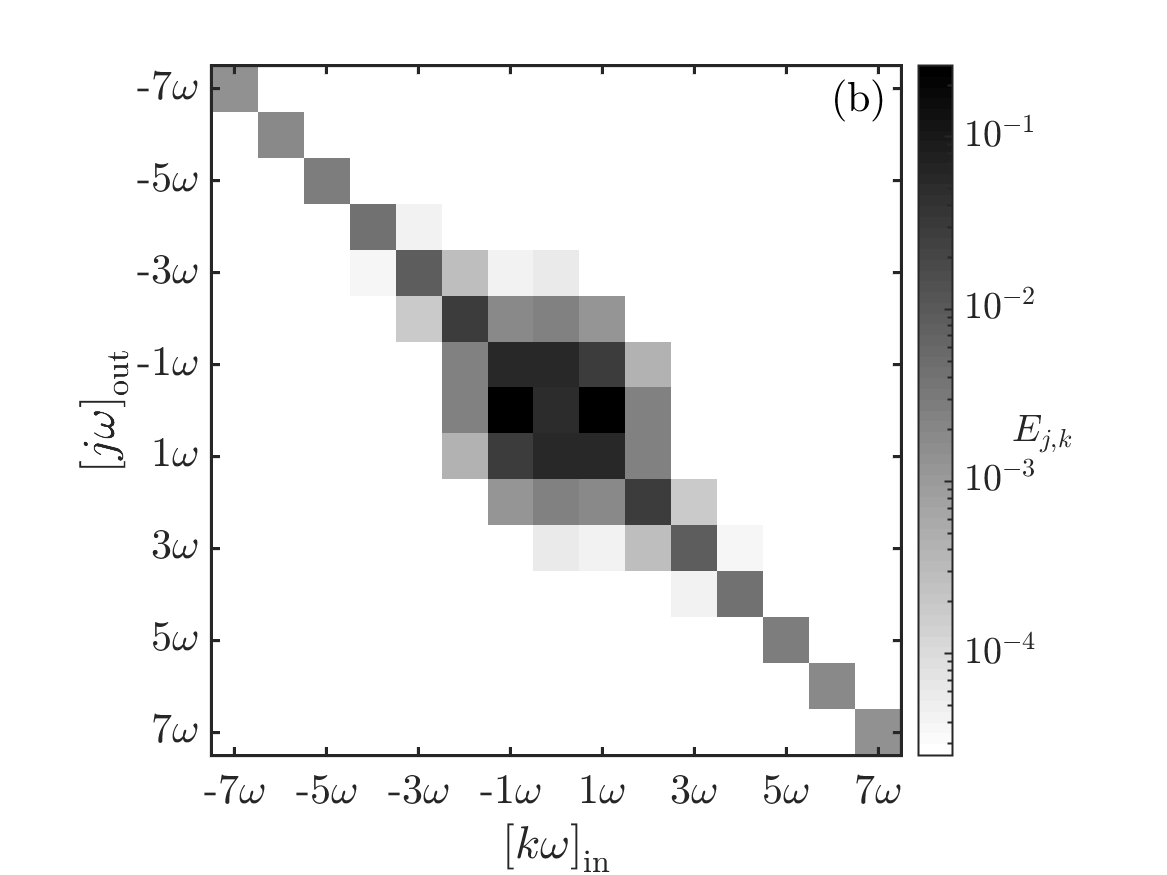}
}
\caption{Fractional variance~\eqref{eqn: fracEnergy} contained within each block of the harmonic
  resolvent for the toy model~\eqref{eqn: 3odes} (a) evaluated about the temporal mean $\Omega_b = \{0\}$, and (b)
  evaluated about the exact periodic solution $\Omega_b = \{k\omega\}_{k\in\{-3,\ldots,3\}}$ (b).}
\label{fig: crossFreqs_maps}
\end{figure}

Figure \ref{fig: response_toyModel} shows the energy spectrum of the
perturbations $\hat\vq'$ in response to the periodic input $\hat\vw'$ as well as
the state evolution $\vq'(t)$ over one fundamental period. We observe from
the nonlinear simulation that forcing at frequency $\omega$ leads to a response
with energy content also at the zeroth, second and third harmonics. We observe
also that the prediction obtained by linearizing about the exact limit cycle
accurately matches the ground truth. This is because the
time-varying base flow about which we evaluate the harmonic resolvent couples structures at
different frequencies and we are therefore able to predict (to first order) the
frequency off-scatter that is observed in the nonlinear system.
The extent to which we are able to capture cross-frequency interactions is given
by the block singular values of~$\Htilde$ shown in figure \ref{fig:
  crossFreqs_maps}b. We color-code the cross-frequency blocks according to the
fractional variance
\begin{equation}
\label{eqn: fracEnergy}
    E_{j,k} = \frac{\sum_{m = 1}^3 \sigma^2_{m,\left(j,k\right)}}{\sum_{n = 1}^{N-1}\sigma_n^2}
\end{equation}
where $\sigma_{m,\left(j,k\right)}$ is the $m^{\mathrm{th}}$ singular value of
the $\left(j\omega,k\omega\right)$ block of $\Htilde$ and $\sigma_n$ is the
$n^{\mathrm{th}}$ singular value of~$\Htilde$.  The normalization is such
that $\sum_{j,k} E_{j,k} = 1$.  We observe from the $1\omega$-column in figure \ref{fig: crossFreqs_maps}b that forcing at the fundamental frequency may trigger a response with spectral energy content up to the third harmonic, and that is precisely what the energy spectrum in figure \ref{fig: response_toyModel}a confirms.  

Linearizing about the temporal mean, however, does not provide an accurate
representation of the response of the nonlinear system to the given periodic
forcing at frequency~$\omega$. First, observe that through this
linearization we overestimate the spectral energy at the fundamental
frequency. Secondly, it is clear from figures \ref{fig: response_toyModel}a and
\ref{fig: response_toyModel}b that the prediction is monochromatic at frequency
$\omega$. This is because~$\Htilde$ is block diagonal, as mentioned at the end of
Section~\ref{sec:incompressible-fluids} and illustrated in
Figure~\ref{fig: crossFreqs_maps}a.  Therefore, no cross-frequency interaction can be
accounted for through the base flow, and forcing at frequency~$\omega$
will only produce a response at the same frequency.


\section{Application to flow past an airfoil at near-stall angle of attack}
\label{sec: airfoilSec}
We now consider two-dimensional incompressible flow past an airfoil at an angle
of attack, under conditions for which there is unsteady vortex shedding.  We
perform numerical simulations using the immersed boundary formulation
of~\cite{ibpm1}, to compute the flow past a NACA~0012 airfoil at
angle of attack of $20^\circ$ and Reynolds number of $200$ based on the chord.  The
immersed boundary formulation enforces no-slip boundary conditions at the
surface $\mathcal{S}$ of the airfoil by imposing a body force $\vf$, as expressed below:
\begin{align}
\label{eqn: immersedBody}
\begin{split}
    \frac{\partial}{\partial t}\vu + \vu\cdot \nabla \vu &= -\nabla p + \Rey^{-1} \nabla^2 \vu + \int_{\mathcal{S}}\vf(\vxi)\delta(\vxi - \vx)\mathrm{d}\vxi\\
    \nabla \cdot \vu &= 0 \\
    \vu(\vxi) &= \int_{\mathcal{X}}\vu(\vx)\delta(\vx -
    \vxi)\mathrm{d}\mathbf{x} = 0,\qquad\text{for $\vxi\in\mathcal{S}$.}
\end{split}
\end{align}
where $\vu(\vx,t)$ and $p(\vx,t)$ are the velocity and pressure over the spatial
domain $\mathcal{X} = \mathbb{R}^2$, and $\delta$ is the Dirac delta function.
The third equation in (\ref{eqn: immersedBody}) is a set of algebraic
constraints that enforce the no-slip boundary condition on the surface~$\mathcal{S}$. We refer to \citet{ibpm1} for a detailed discussion of the method.

\begin{figure}
\centering
\subfloat
{
\includegraphics[trim = 0.37in 0in 1.3in 0.1in, clip,width=0.4\textwidth]{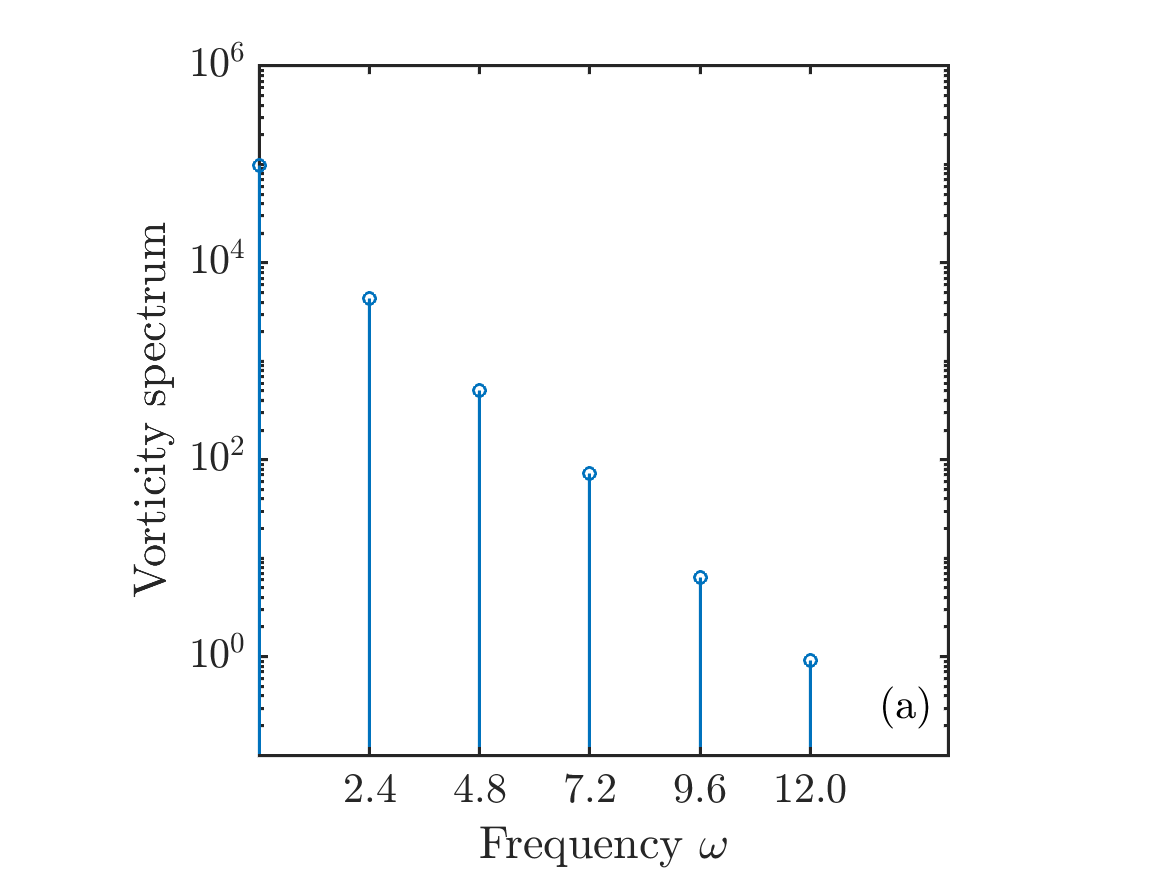}
}
\hspace{1ex}
\subfloat
{
\includegraphics[trim = 0in 0.1in 0.4in 0.1in, clip,width=0.55\textwidth]{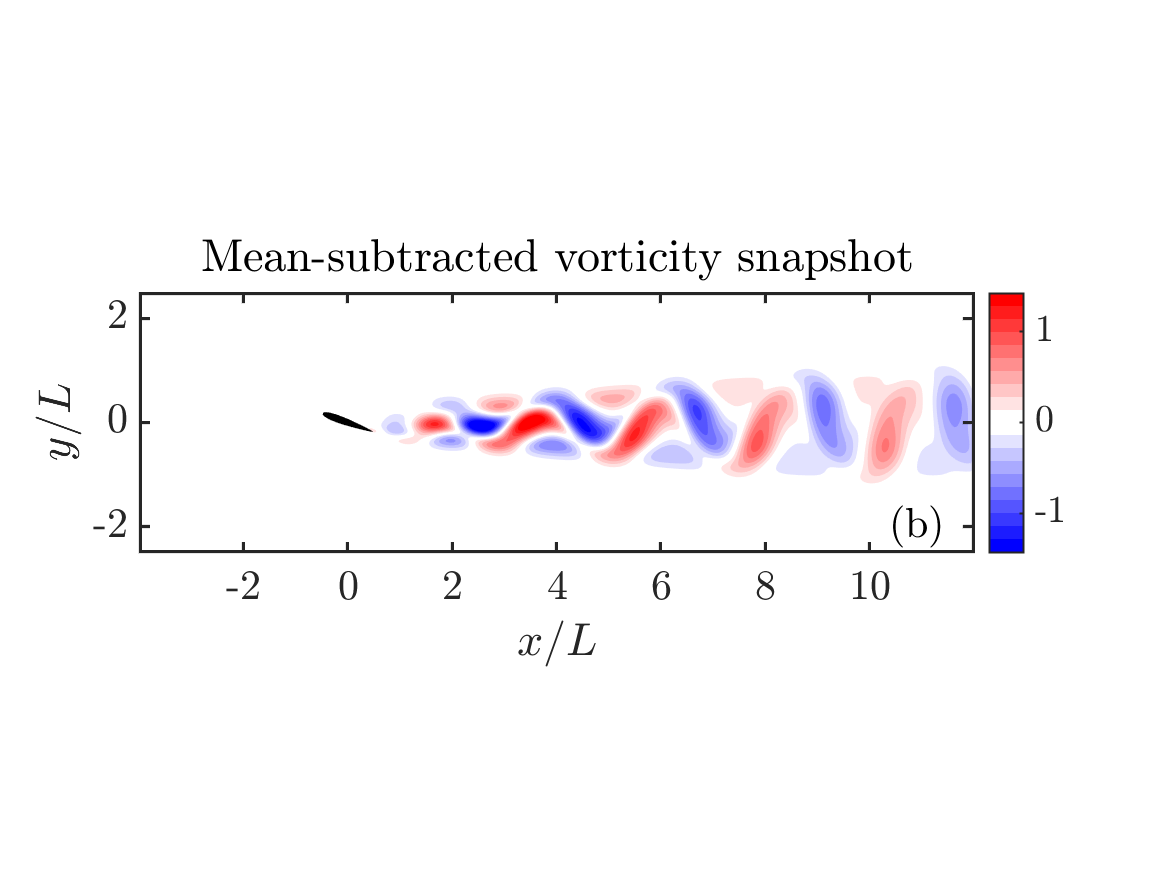}
}
\caption{Nonlinear simulation of flow past an airfoil, showing (a) vorticity spectrum;
    and (b) mean-subtracted vorticity snapshot on the limit cycle.}
\label{fig: airfoilSnap}
\end{figure}

We center the half-chord of the airfoil at the origin of the computational
domain of size $\left[-4,12\right]\times\left[-2.5,2.5\right]$ and we discretize
the domain on a $800\times 250$ grid. We impose a uniform inflow boundary condition at the inlet, slip-wall boundary conditions at the top and bottom boundaries and a convective outflow boundary condition at the outlet. The vorticity spectrum is shown in figure \ref{fig: airfoilSnap}a, while a representative snapshot of the mean-subtracted vorticity field on the limit cycle is shown in figure \ref{fig: airfoilSnap}b. We observe that up to five harmonics of the fundamental frequency $\omega = 2.40$ are active on the limit cycle, suggesting that non-trivial nonlinear mechanisms are at play. 

In the upcoming analysis we take our state vector to be $\vq=(\vu,p,\vf)$
and we expand the dynamics about a chosen base flow $\vQ(t)$. We omit the spatial dependence of the states for notational simplicity. Moreover we consider perturbations $\vq'(t)$ over the set of frequencies $\Omega = \{-7\omega,\ldots,7\omega \}$. Upon linearizing the dynamics about the chosen base flow we obtain the linear input-output system 

\begin{equation}
    \hat\vq' = \Htilde\hat\vw' = \sum_{j = 1}^{N-1} \sigma_j\hat\vphi_{j}\hat\vpsi^*_{j}\hat\vw'
\end{equation}
where $\hat\vw'$ is the frequency-domain representation of the nonlinear terms that feed back into the linear harmonic resolvent. The left singular vector $\hat\vphi_j$ is the $j^{\text{th}}$ global
output mode and the right singular vector $\hat\vpsi_j$ is the $j^{\text{th}}$
global input mode. 
We wish to specify that
$\Htilde$ is not computed explicitly since it is a dense operator of
prohibitive size $N \sim O\left(10^7 \right)$. Specifically, given the $n$-dimensional state vector $\vq'$ and $15$ frequencies in $\Omega$, the size of the harmonic resolvent is $N = 15n$.
Instead, given $\mH^{-1}$, which is a sparse operator whose nonzero entries depend on the spatial
discretization scheme used on the governing equations, we computed the leading
singular values and singular vectors of $\Htilde$ using one of the randomized
singular value decomposition algorithms in \citet{rsvd}. The implementation was
carried out with an in-house solver based on the PETSc (\cite{petsc}) and SLEPc (\cite{slepc}) libraries.

\subsection{Amplification mechanisms about a time-varying base flow}
We linearize the dynamics in (\ref{eqn: immersedBody}) about a time-periodic base flow over
the set of frequencies $\Omega_b =\{-3\omega,\ldots,3\omega\}$ with
$\omega = 2.40$ as in figure~\ref{fig: airfoilSnap}a, and we compute the
singular value decomposition of the harmonic resolvent.

\begin{figure}
\centering
\subfloat
{
\includegraphics[trim = 0.5in 0in 0.6in 0.1in, clip,width=0.45\textwidth]{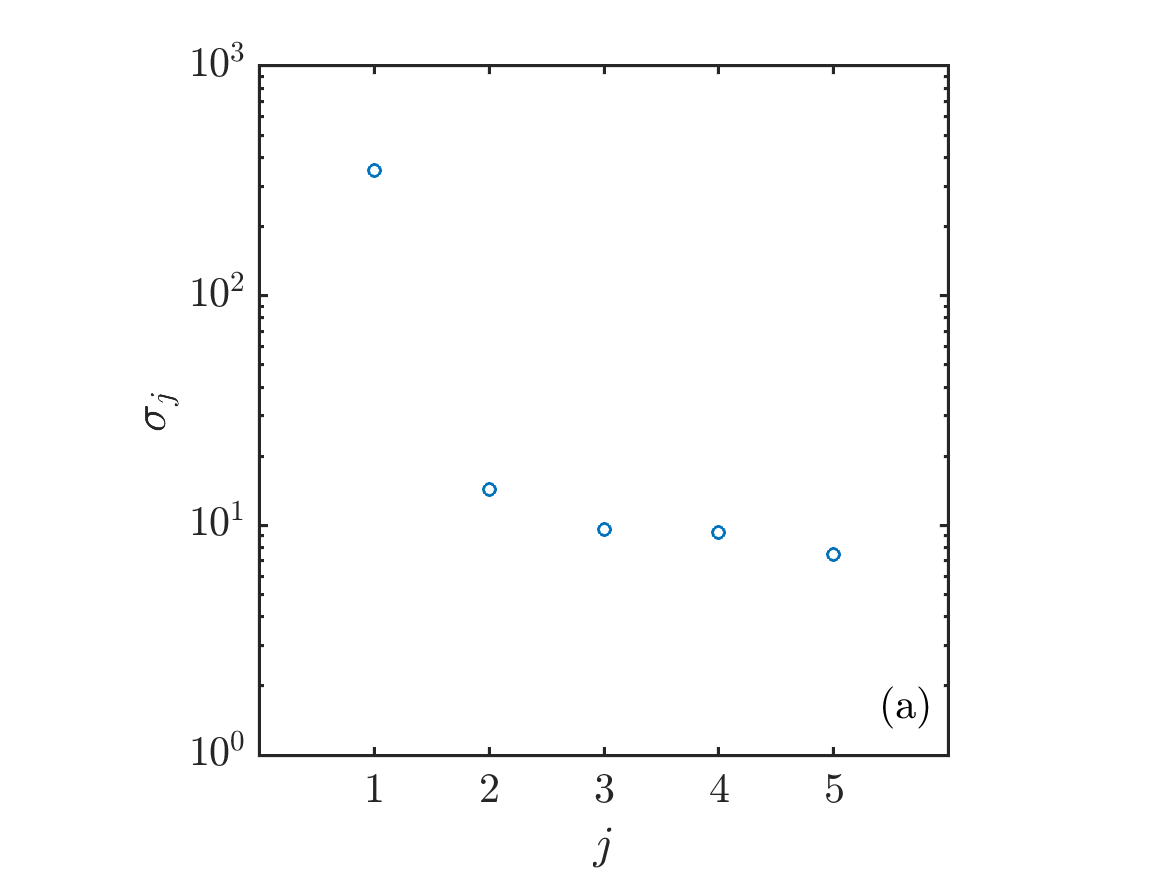}
}
\hspace{1ex}
\subfloat
{
\includegraphics[trim = 0.5in 0in 0.6in 0.1in, clip,width=0.45\textwidth]{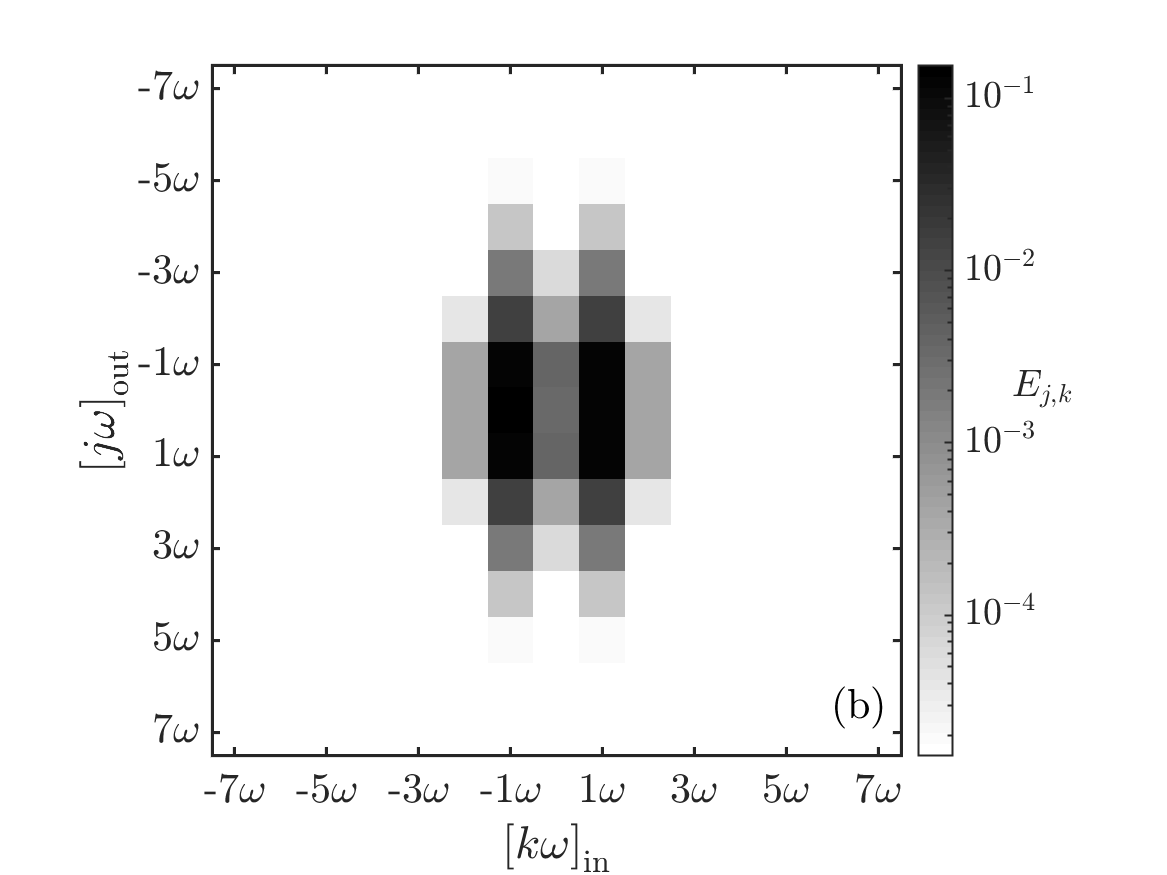}
}
\caption{Singular values of the harmonic resolvent for flow past an airfoil with $\Omega_b = \{-3\omega,\ldots,3\omega\}$, $\Omega = \{-7\omega,\ldots,7\omega\}$ and $\omega = 2.40$, showing (a) singular values of $\Htilde$, and (b) block-wise fractional variance $E_{j,k}$ defined by an expression similar to (\ref{eqn: fracEnergy}).}
\label{fig: htfAmpl}
\end{figure}

First, note from figure~\ref{fig: htfAmpl}a that there is more than an
order of magnitude separation between the first and the second singular values
of the harmonic resolvent and we can therefore argue that it has low-rank structure.

Second, figure~\ref{fig: htfAmpl}b shows that the nonlinear flow is very
susceptible to perturbations at the fundamental frequency, since the block
singular values of~$\Htilde$ suggest that introducing forcing at $\omega$ has an
effect on flow structures up to the $4^{\mathrm{th}}$ harmonic. Likewise, we can
conclude that the flow is less sensitive to perturbations at higher harmonics of
the fundamental frequency as we observe that the singular values of the blocks
governing those dynamics are one (or more) orders of magnitude less than those in the $1\omega$-column.

\begin{figure}
\centering
\subfloat
{
\includegraphics[trim = 0.25in 1.1in 0.2in 1.1in, clip,width=0.44\textwidth]{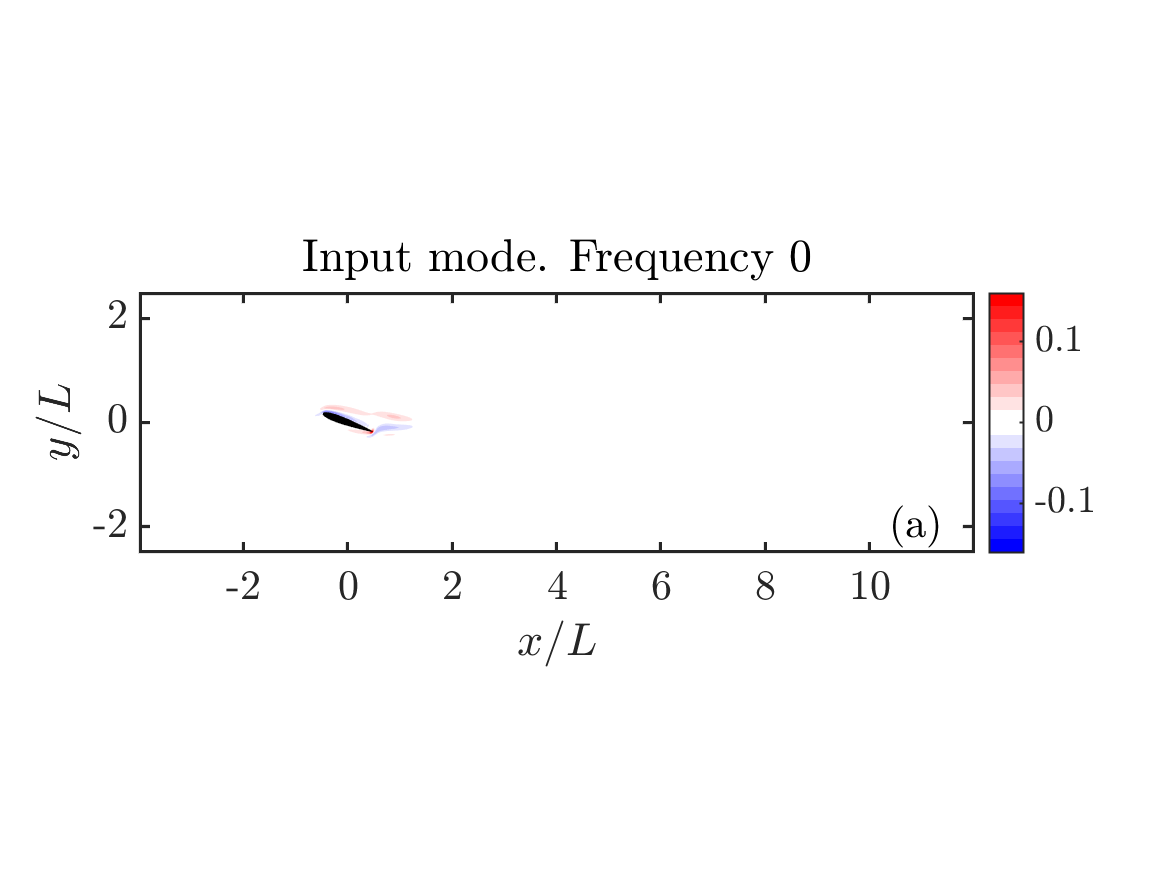}
}
\hspace{1ex}
\subfloat
{
\includegraphics[trim = 0.25in 1.1in 0.2in 1.1in, clip,width=0.44\textwidth]{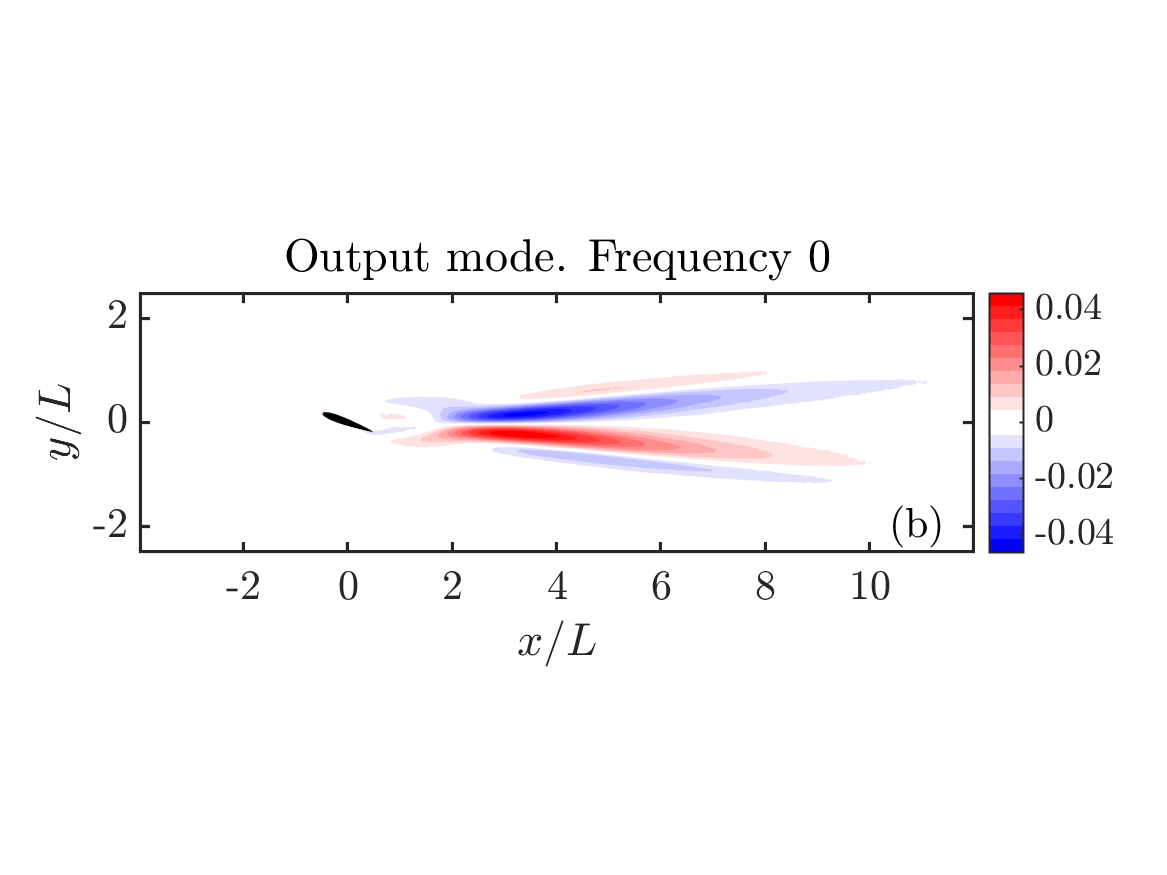}
}

\subfloat
{
\includegraphics[trim = 0.25in 1.1in 0.2in 1.1in, clip,width=0.44\textwidth]{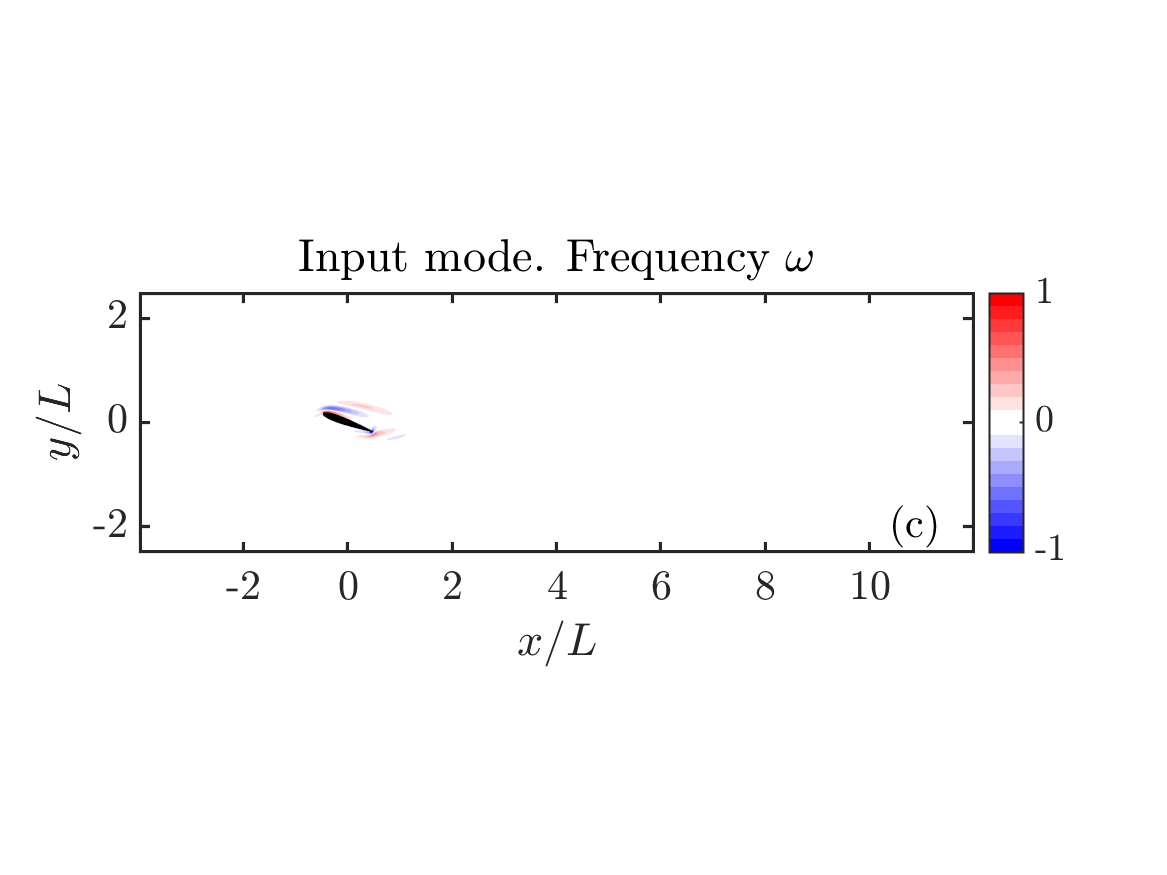}
}
\hspace{1ex}
\subfloat
{
\includegraphics[trim = 0.25in 1.1in 0.2in 1.1in, clip,width=0.44\textwidth]{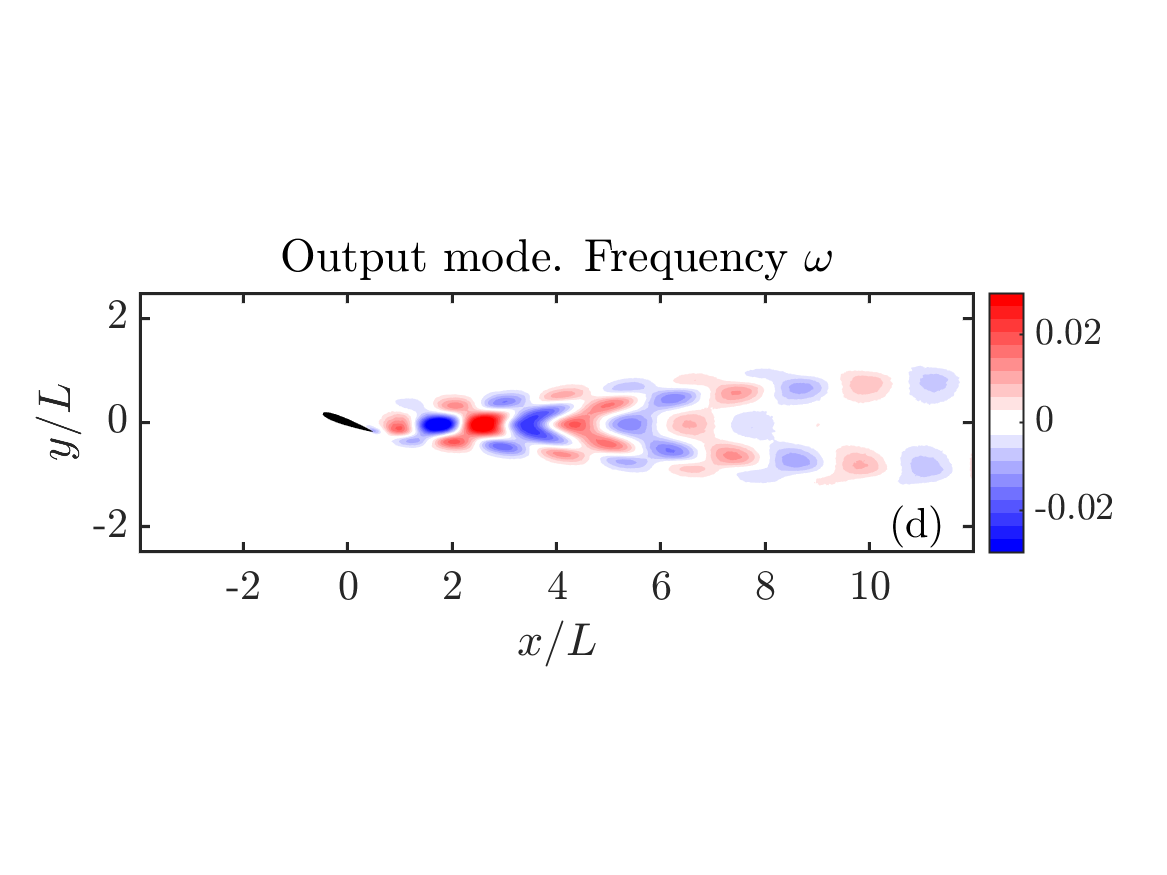}
}

\subfloat
{
\includegraphics[trim = 0.25in 1.1in 0.2in 1.1in, clip,width=0.44\textwidth]{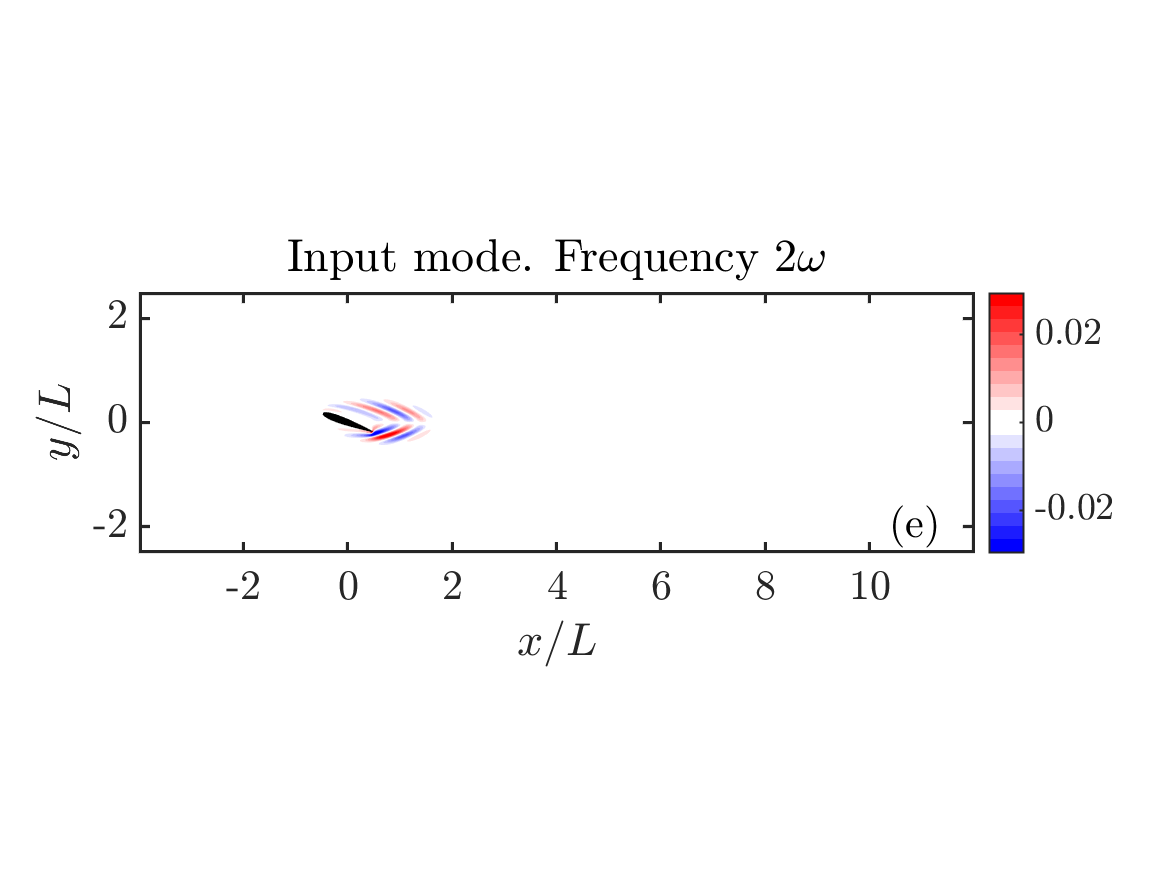}
}
\hspace{1ex}
\subfloat
{
\includegraphics[trim = 0.25in 1.1in 0.2in 1.1in, clip,width=0.44\textwidth]{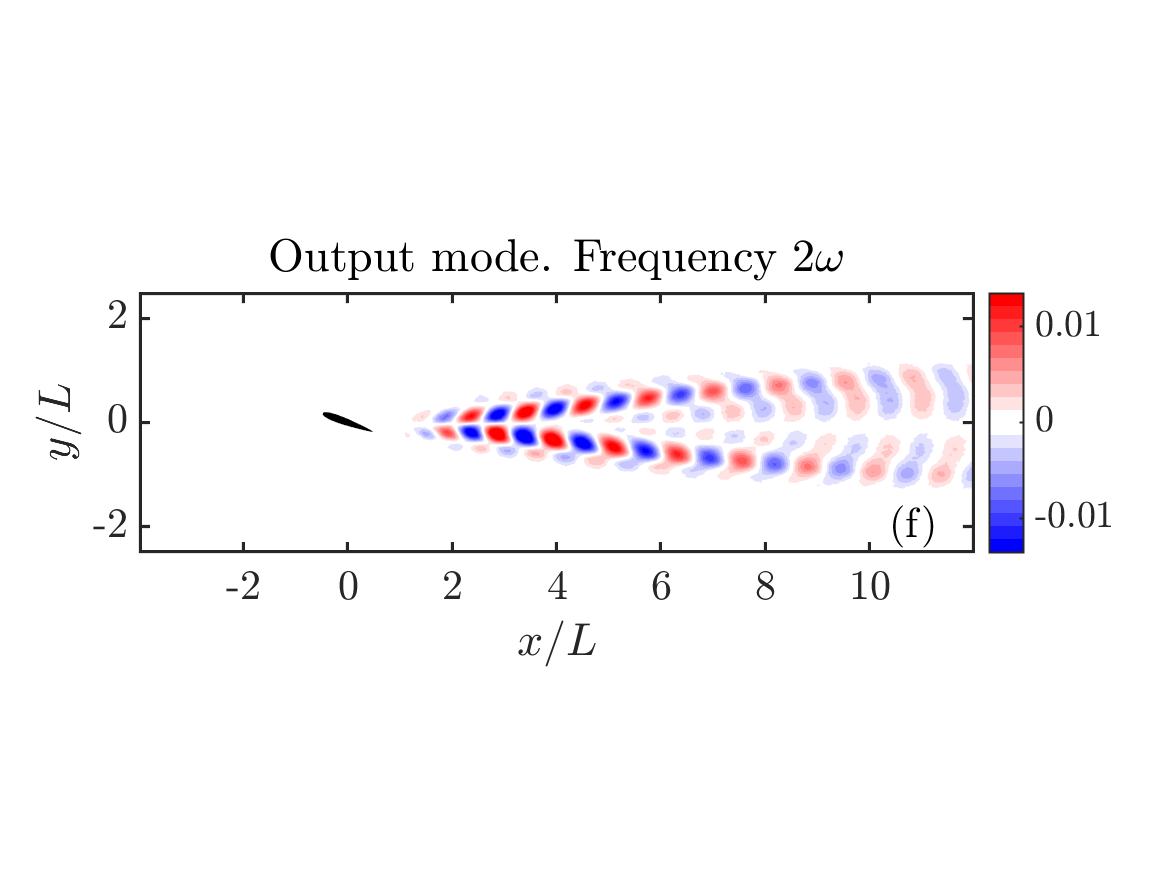}
}

\subfloat
{
\includegraphics[trim = 0.25in 1.1in 0.2in 1.1in, clip,width=0.44\textwidth]{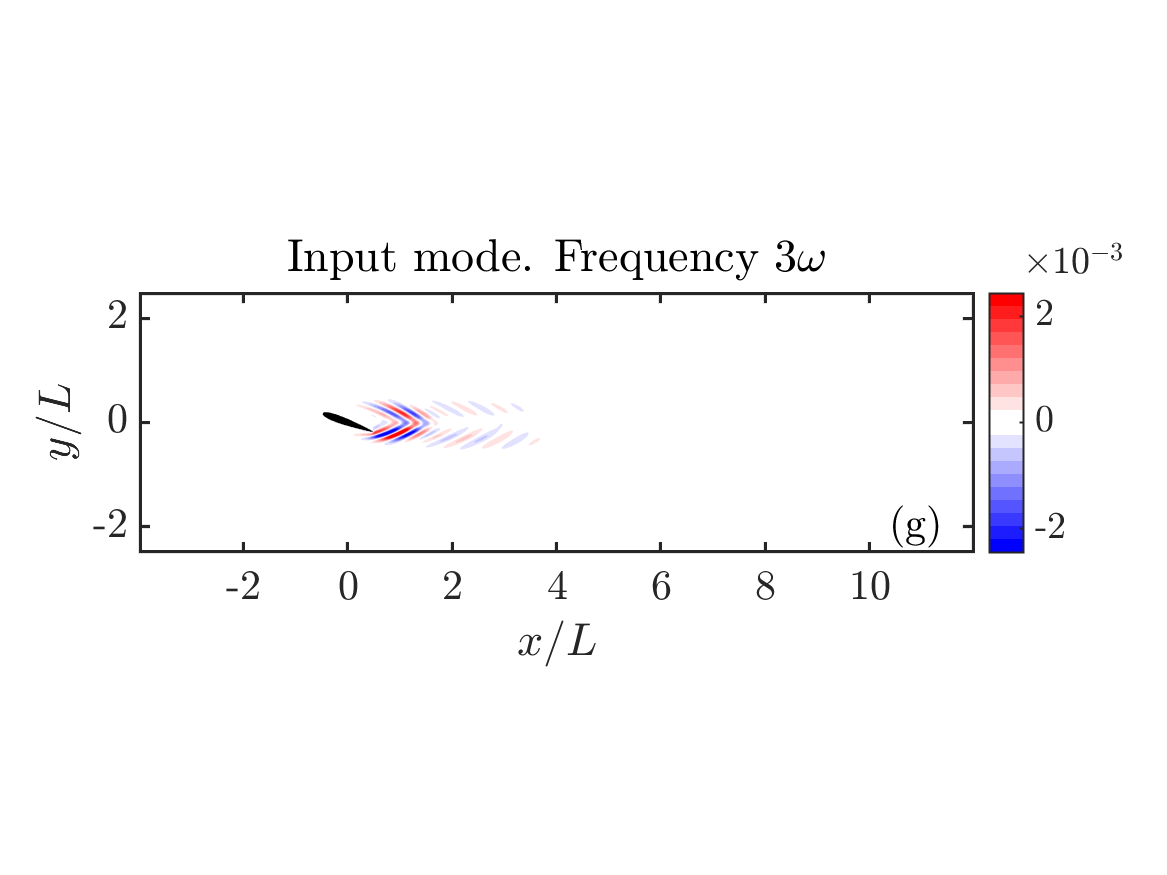}
}
\hspace{1ex}
\subfloat
{
\includegraphics[trim = 0.25in 1.1in 0.2in 1.1in, clip,width=0.44\textwidth]{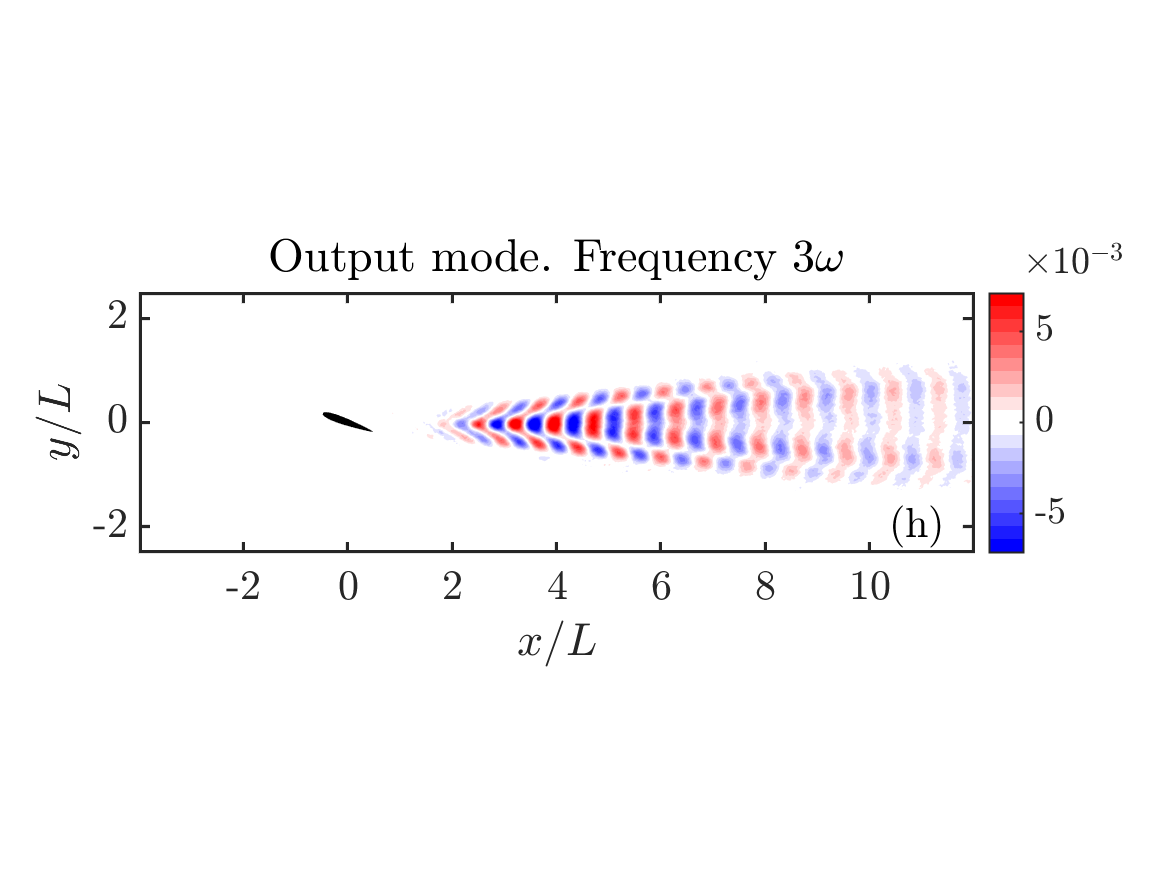}
}

\caption{Real part of the vorticity field computed from the input mode and
  the output mode associated with $\sigma_1$ in figure~\ref{fig: htfAmpl}a.}
\label{fig: htfModes}
\end{figure}

We may also draw conclusions about the sensitivity of the flow from the
$k\omega$-entries of the first input mode of~$\Htilde$, shown in
figure~\ref{fig: htfModes}. Recall from the previous sections that the input
modes describe the spatio-temporal structures that are most effective at
exciting a response, while the output modes describe the spatio-temporal
structures that are preferentially excited by these inputs. Specifically, we
learn from the magnitude of the entries of the input mode that the flow is most
sensitive to perturbations at frequency $\omega$ and it is the least sensitive
to perturbations at frequency $3\omega$. Moreover, the mode shapes of the
entries of the input mode suggest that the flow is very sensitive to
perturbations that are spatially localized around the body, while the output
mode entries in figures~\ref{fig: htfModes} (b), (d), (f) and~(h) illustrate the
spatial structures that should arise in the flow in response to a disturbance or
control input that aligns well with the input mode. In order to verify this
statement, we introduce a small amplitude forcing in the flow by sinusoidally
moving the airfoil in the vertical direction, with velocity
\begin{equation}
  \label{eq:11}
  v = \varepsilon\sin\omega t.
\end{equation}
The forcing frequency is taken to be the fundamental frequency of vortex
shedding, $\omega = 2.40$, while $\varepsilon = 0.01$. 

In order to draw a direct comparison with the output modes of the harmonic
resolvent operator we proceed as follows. We let the flow reach the limit cycle
$\vQ(t)$ whose spectrum is shown in figure \ref{fig: airfoilSnap}a and then we
introduce the sinusoidal motion described in~\eqref{eq:11}.  We let the flow
evolve for a few periods until transients have died out and then we subtract the
base flow $\vQ(t)$, so that
\begin{equation}
    \vq'_f(t) = \vq_f(t) - \vQ(t),
  \end{equation}
  where $\vq_f$ denotes the state of the forced flow.
It is worth mentioning that the forcing in the nonlinear simulation may
introduce a phase shift relative to the limit cycle, and it might therefore be
necessary to phase match $\vq_f(t)$ and $\vQ(t)$ before computing
$\vq'_f(t)$. Finally, we Fourier transform $\vq'_f(t)$ and we compare the
resulting modes to the the output modes of the harmonic resolvent. 

\begin{figure}
\centering
\subfloat
{
\includegraphics[trim = 0.25in 1.1in 0.2in 1.1in, clip,width=0.44\textwidth]{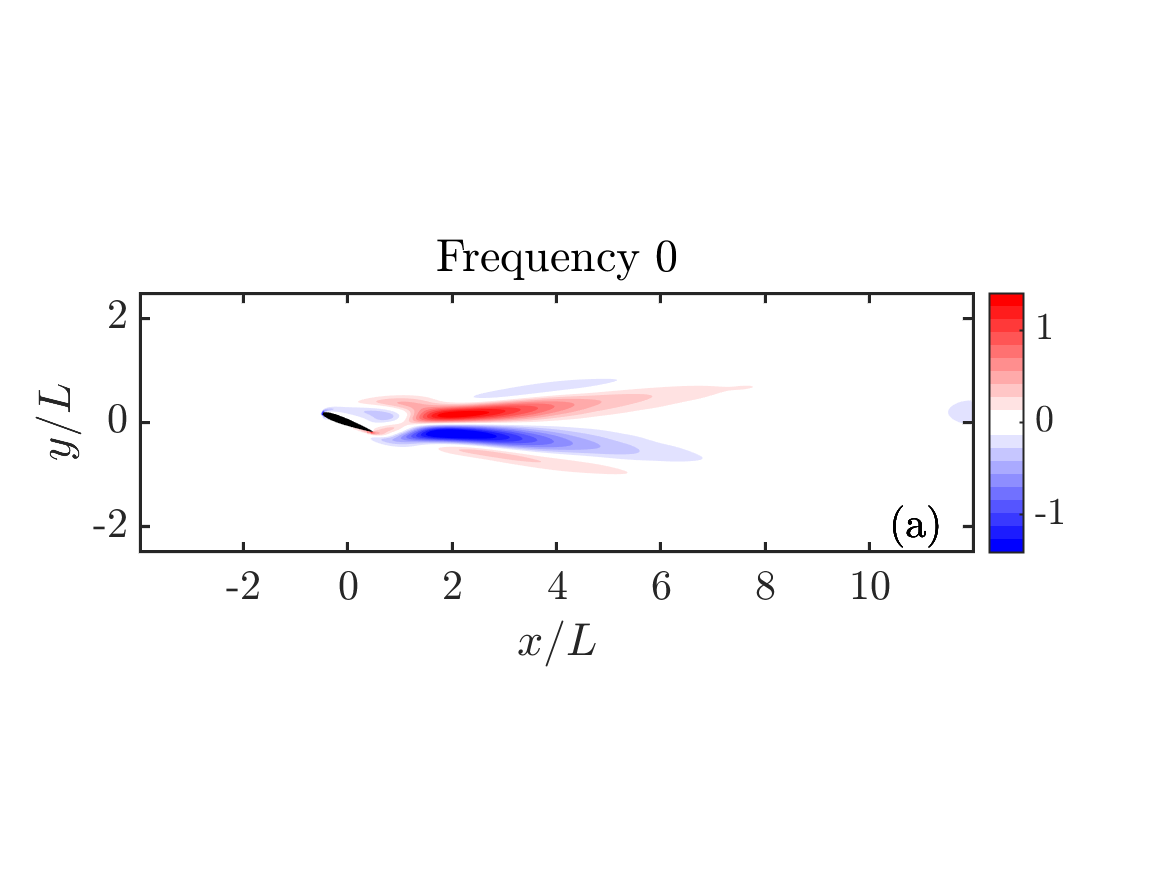}
}
\hspace{1ex}
\subfloat
{
\includegraphics[trim = 0.25in 1.1in 0.2in 1.1in, clip,width=0.44\textwidth]{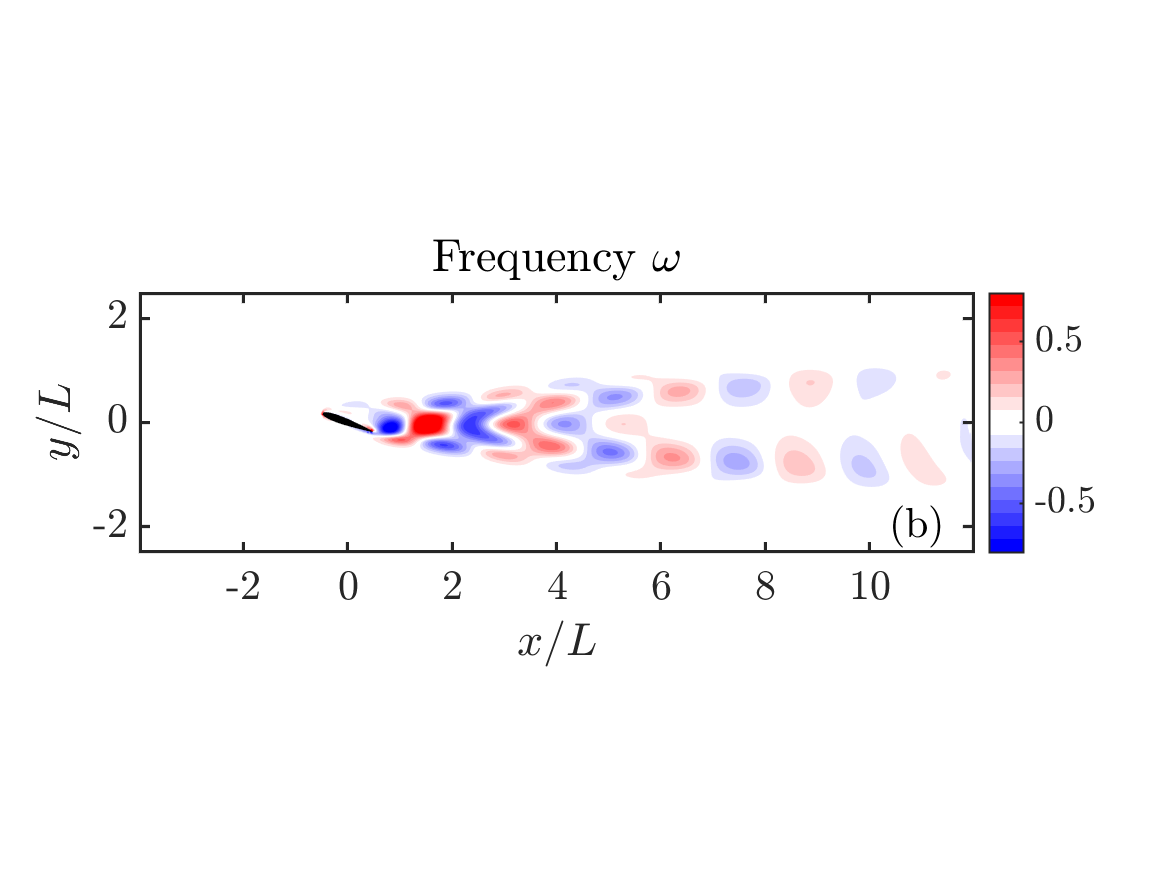}
}

\subfloat
{
\includegraphics[trim = 0.25in 1.1in 0.2in 1.1in, clip,width=0.44\textwidth]{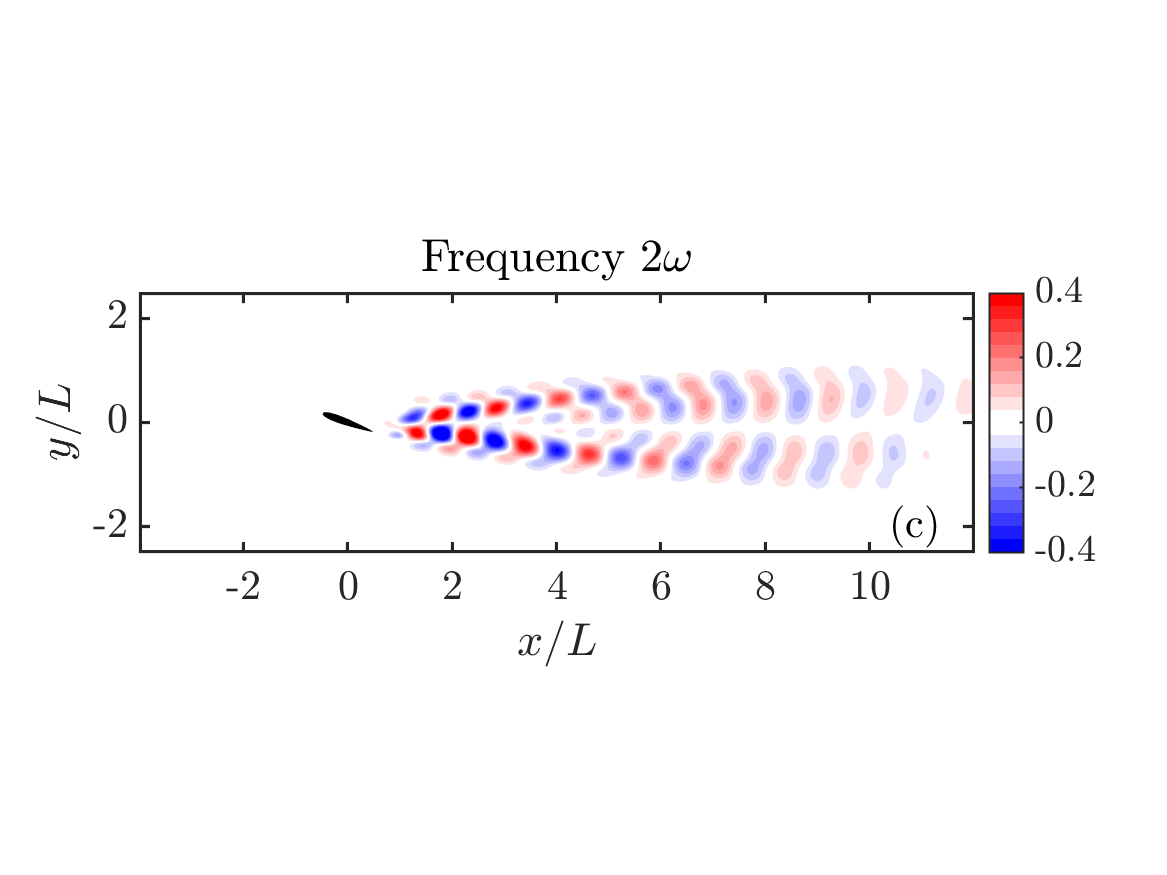}
}
\hspace{1ex}
\subfloat
{
\includegraphics[trim = 0.25in 1.1in 0.2in 1.1in, clip,width=0.44\textwidth]{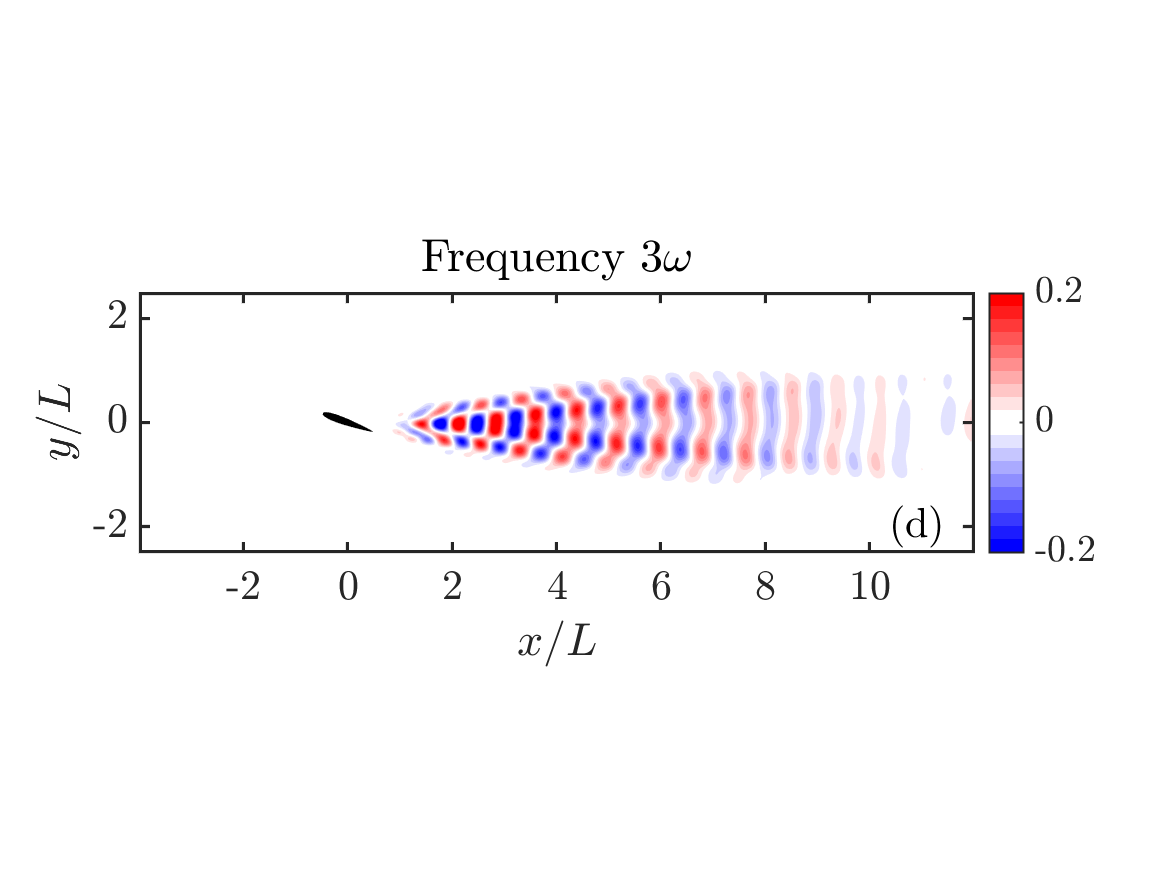}
}

\caption{Real part of the Fourier modes of the vorticity perturbations for the airfoil
with sinusoidal motion~\eqref{eq:11} for frequencies $0$ through $3\omega$.}
  \label{fig: fourier_perturbations}
\end{figure}

Figure~\ref{fig: fourier_perturbations} shows the vorticity of the first few
Fourier modes computed from $\vq'_f(t)$.  These highlight the vortical
structures that result when the flow is forced sinusoidally according
to~\eqref{eq:11}. Remarkably, the first output mode of the harmonic resolvent,
shown in Figure~\ref{fig: htfModes}, provides a surprisingly accurate prediction
of the structures, at all four frequencies shown in Figure~\ref{fig:
  fourier_perturbations}.
This close agreement is presumably a consequence of the low-rank structure of
the harmonic resolvent (see Figure~\ref{fig: htfAmpl}): regardless of the type
of forcing, the resulting flow perturbations will resemble the output modes
shown in Figure~\ref{fig: htfModes}.
The fact that the input modes shown in Figure~\ref{fig: htfModes} are supported
near the airfoil suggests that the flow is sensitive to perturbations near the
airfoil.  While this result is not surprising, the simulations with sinusoidal motion
of the airfoil confirm this behavior.

\subsection{Amplification mechanisms about the temporal mean}
\begin{figure}
\centering
\subfloat
{
\includegraphics[trim = 0.5in 0in 0.6in 0.1in, clip,width=0.45\textwidth]{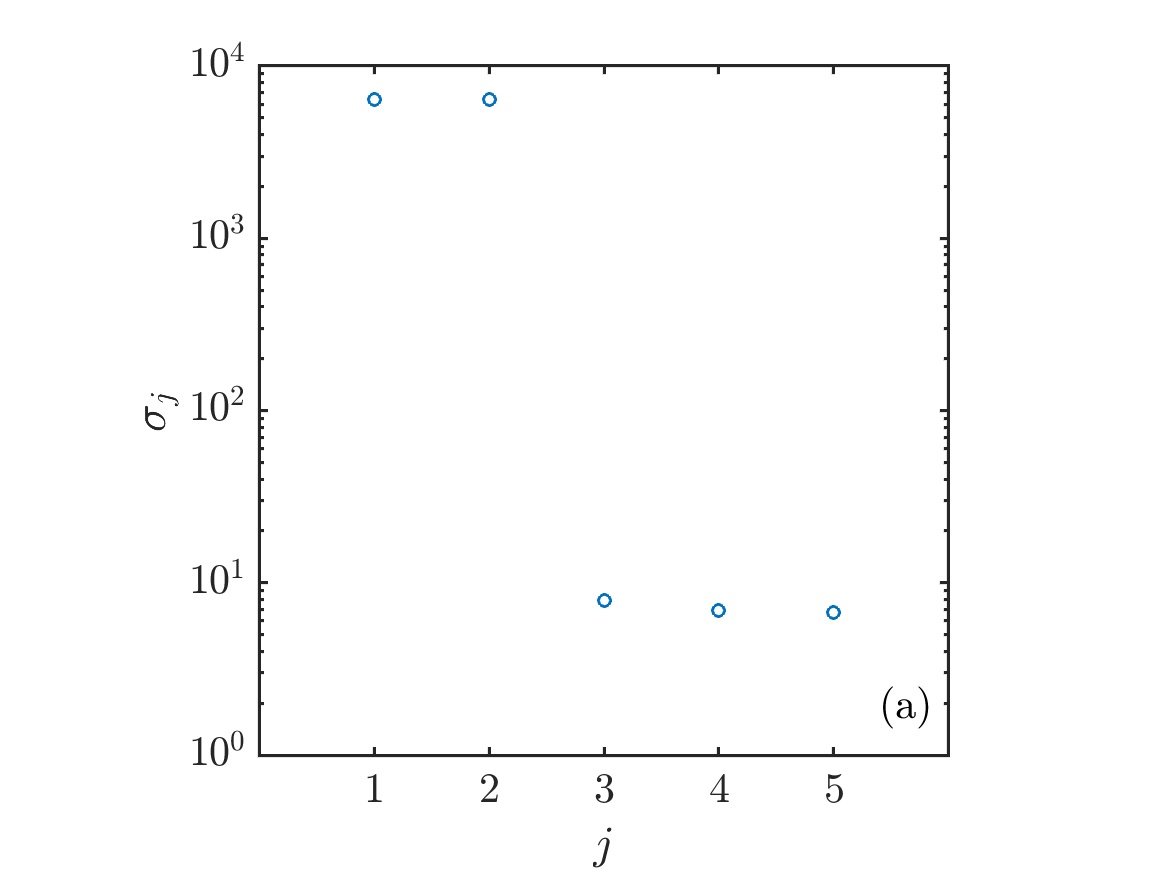}
}
\hspace{1ex}
\subfloat
{
\includegraphics[trim = 0.5in 0in 0.6in 0.1in, clip,width=0.45\textwidth]{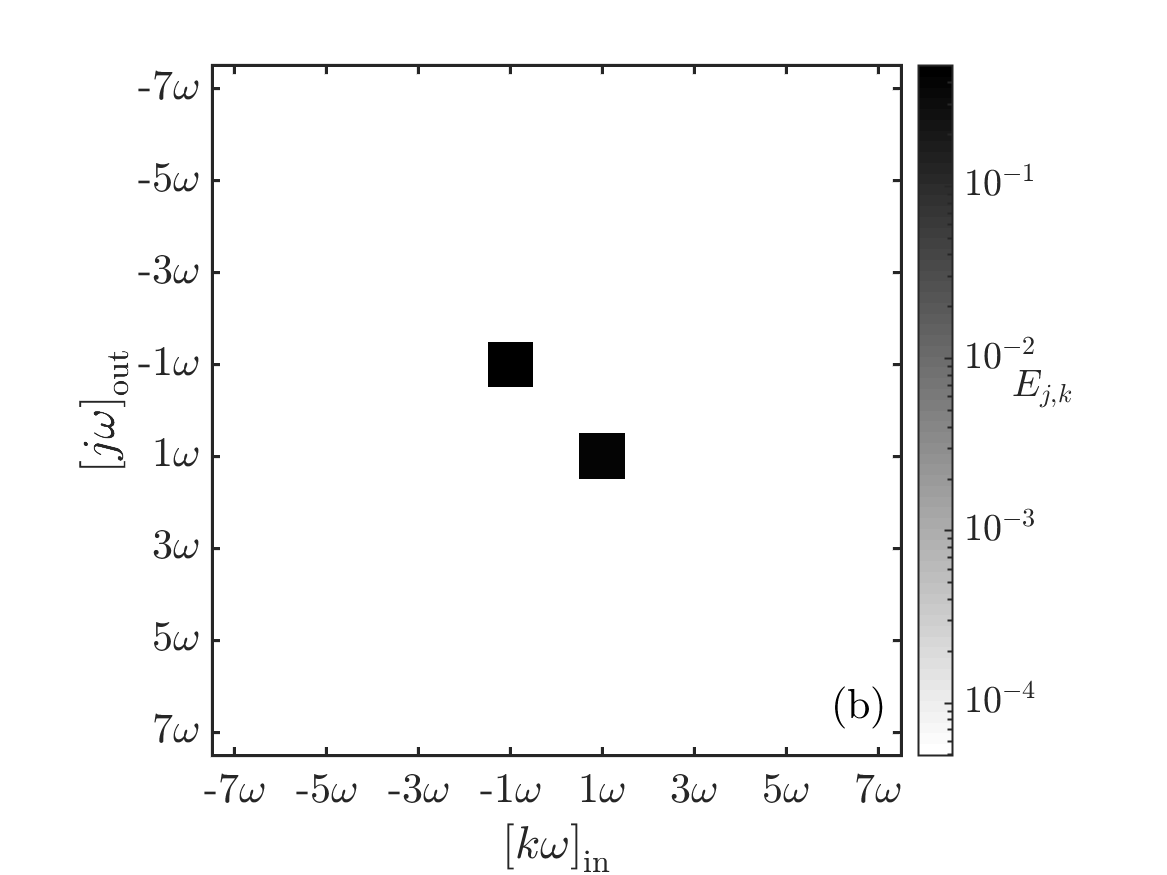}
}
\caption{Singular values for the harmonic resolvent of flow past an airfoil, with $\Omega_b = \{0\}$, $\Omega = \{-7\omega,\ldots,7\omega\}$ and $\omega = 2.40$, showing (a) singular values of $\Htilde$, and (b) block-wise fractional variance $E_{j,k}$ defined by an expression similar to (\ref{eqn: fracEnergy}).}
\label{fig: resAmpl}
\end{figure}

\begin{figure}
\centering
\subfloat
{
\includegraphics[trim = 0.25in 1.1in 0.2in 1.1in, clip,width=0.44\textwidth]{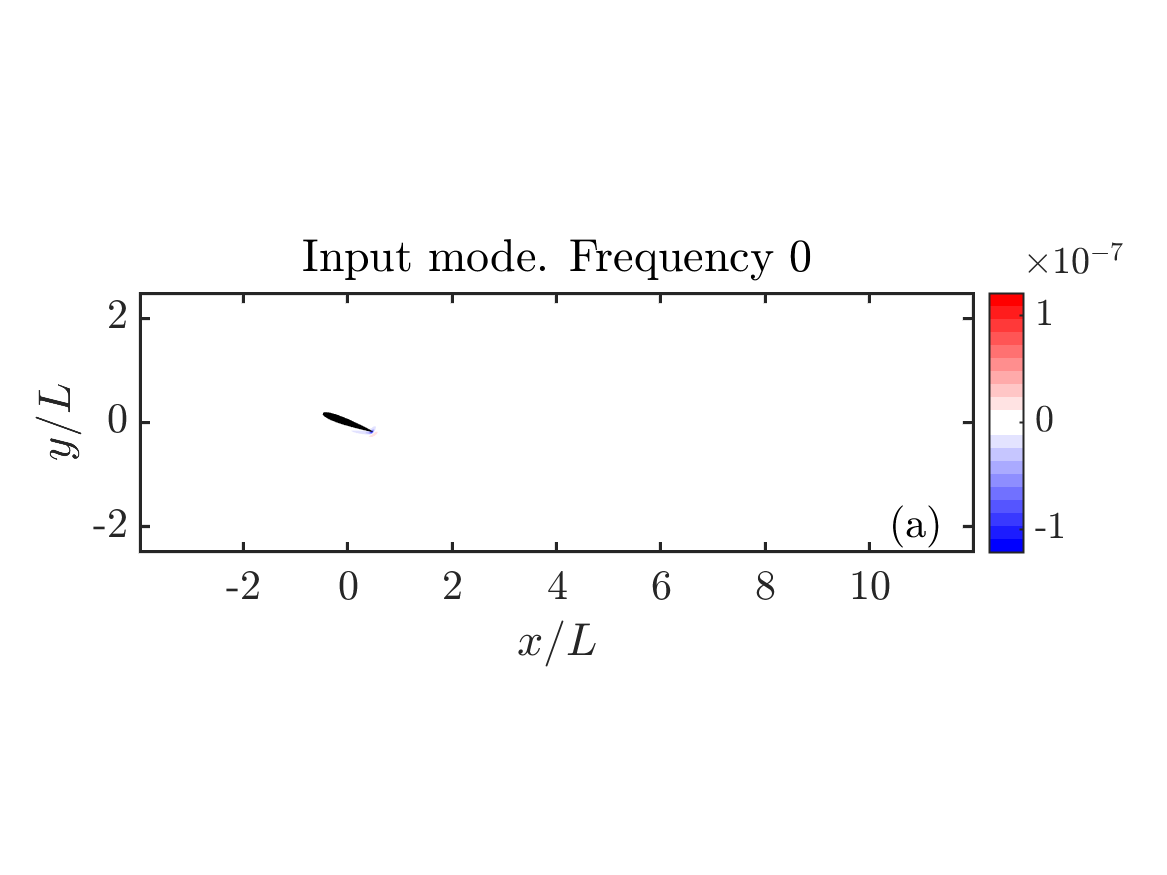}
}
\hspace{1ex}
\subfloat
{
\includegraphics[trim = 0.25in 1.1in 0.2in 1.1in, clip,width=0.44\textwidth]{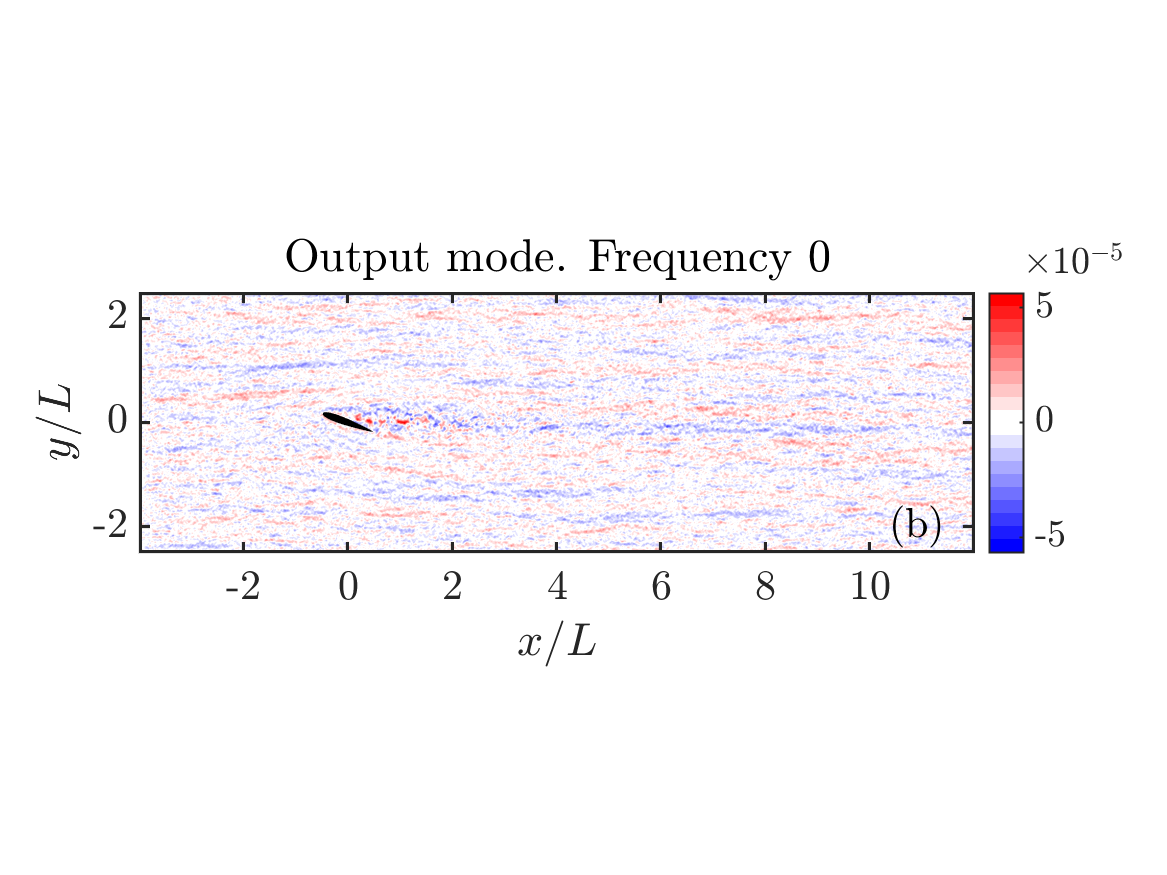}
}

\subfloat
{
\includegraphics[trim = 0.25in 1.1in 0.2in 1.1in, clip,width=0.44\textwidth]{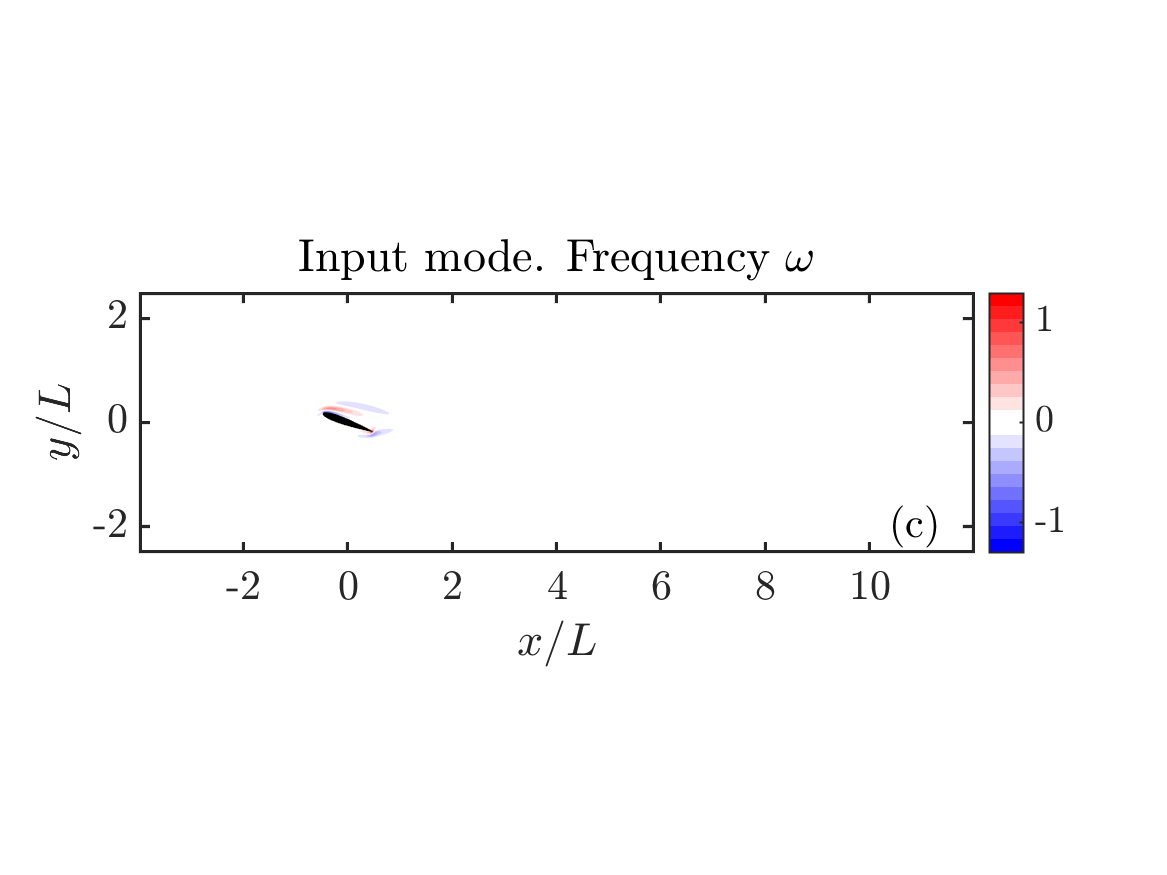}
}
\hspace{1ex}
\subfloat
{
\includegraphics[trim = 0.25in 1.1in 0.2in 1.1in, clip,width=0.44\textwidth]{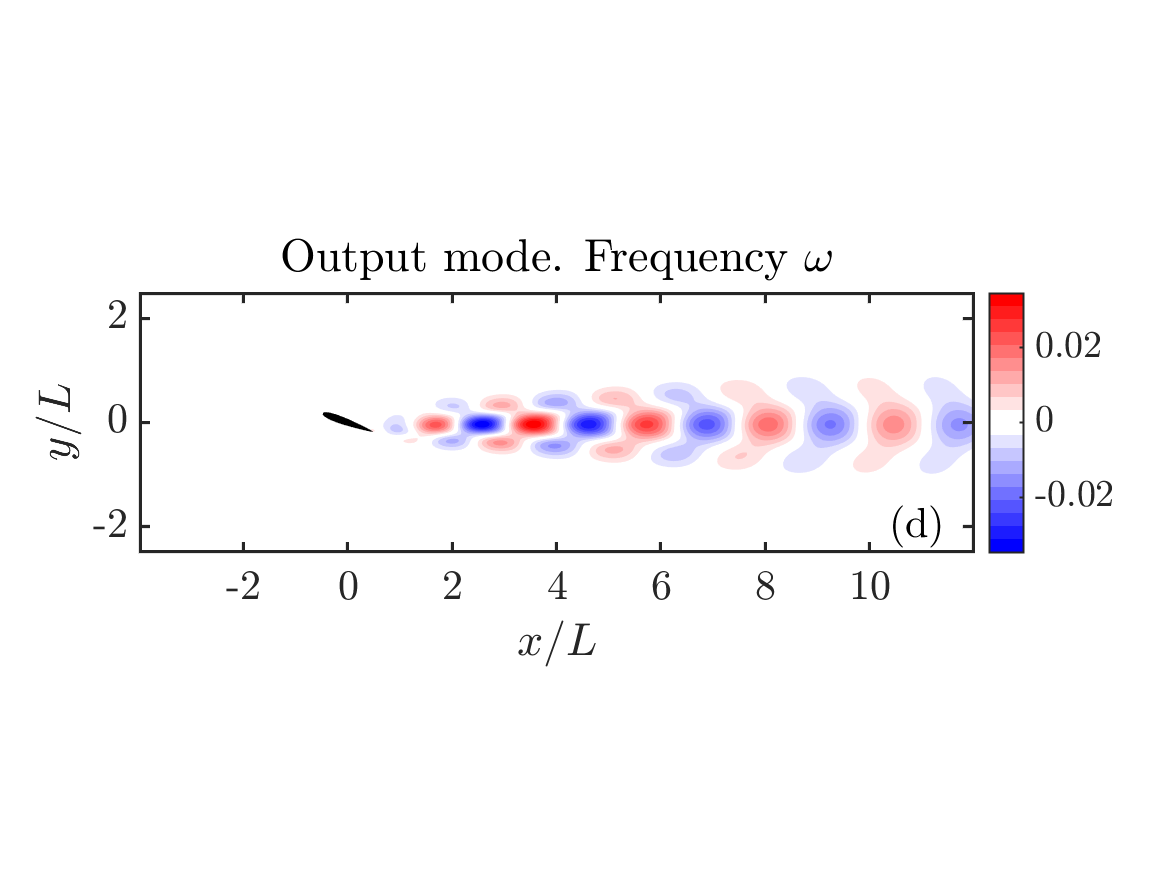}
}

\subfloat
{
\includegraphics[trim = 0.25in 1.1in 0.2in 1.1in, clip,width=0.44\textwidth]{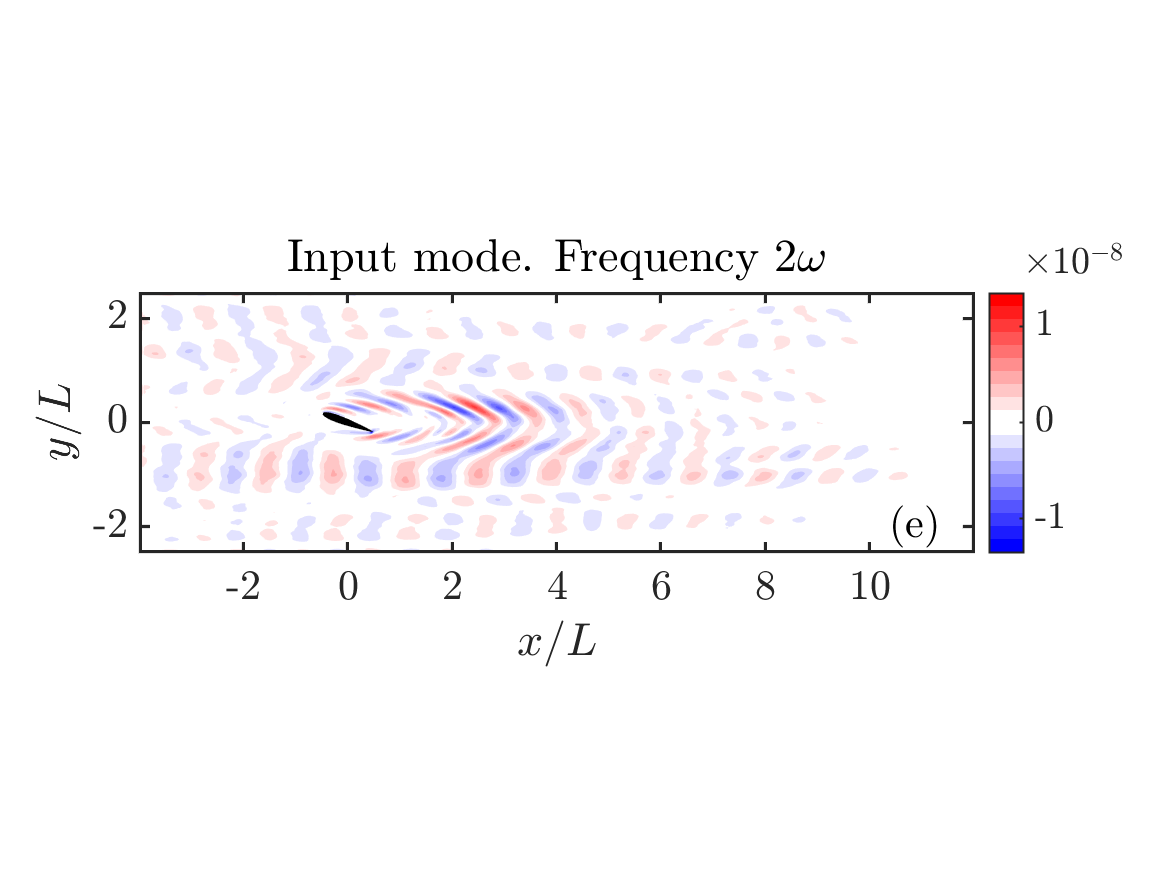}
}
\hspace{1ex}
\subfloat
{
\includegraphics[trim = 0.25in 1.1in 0.2in 1.1in, clip,width=0.44\textwidth]{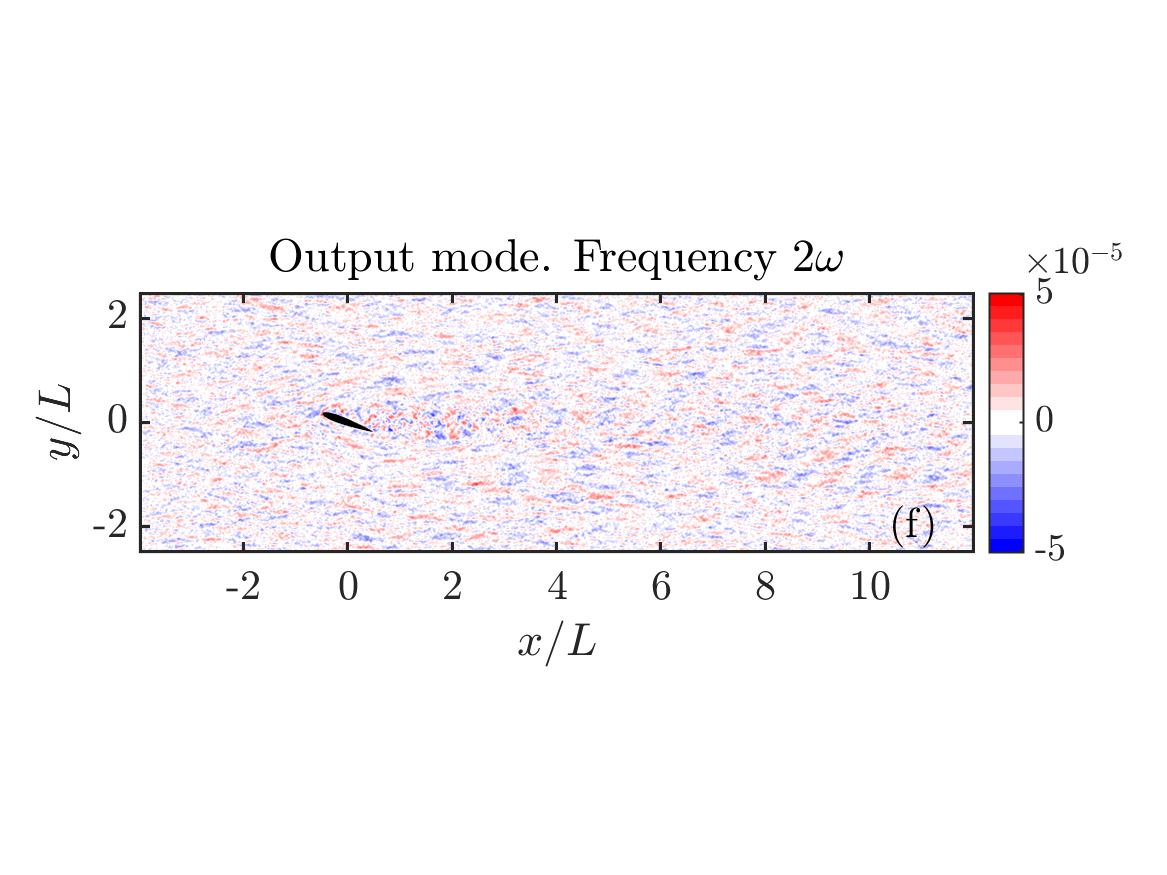}
}

\caption{Real part of the vorticity field computed from the first input
  and output pair of the harmonic resolvent evaluated at the temporal
  mean.}
\label{fig: resolventModes}
\end{figure}

We now evaluate the harmonic resolvent about the temporal mean and compute the
input and output modes as we have done in the previous section. We recall that
in this case, with $\Omega_b = \{0\}$, the harmonic resolvent becomes block
diagonal, and the standard resolvent framework is recovered. Once again, we
consider perturbations over the set of frequencies
$\Omega = \{-7\omega,\ldots,7\omega\}$. The harmonic resolvent was found to be
rank~2, with $\sigma_1 = \sigma_2 \approx 6\times 10^3$, where $\sigma_1$ is the
singular value associated with the resolvent operator at frequency $\omega$,
while $\sigma_2$ is the singular value associated with the resolvent operator at
frequency $-\omega$. The first five singular values are shown in figure
\ref{fig: resAmpl}a. In figure \ref{fig: resolventModes} we show the entries of
the first input and output pair of the harmonic resolvent evaluated at the
temporal mean.

It appears that meaningful information is obtained only at the fundamental
frequency~$\omega$ in figure~\ref{fig: resolventModes}. Specifically, the
$\omega$-component of the input mode provides accurate information on the
sensitivity of the flow, since it is analogous to the $\omega$-component of the
input mode of the harmonic resolvent operator evaluated about the time-varying base flow in figure
\ref{fig: htfModes}c. The corresponding output mode in figure~\ref{fig: resolventModes}, however, does not capture the qualitative behavior
that has been observed in the forced nonlinear simulation, shown in
Figure~\ref{fig: fourier_perturbations}.

Finally, it appears that no meaningful information is provided
by the resolvent operators at the zero frequency or at higher harmonics of the
fundamental frequency~$\omega$.  The reason behind this can be understood by
looking at the diagonal blocks of figures~\ref{fig: resAmpl}b and~\ref{fig:
  htfAmpl}b. Both, in fact, suggest that the temporal mean (block-diagonal
entries of the HTF) does not amplify disturbance at higher frequencies, meaning
that we cannot expect the resolvent operators at frequencies $k\omega$ with
$k\neq 1$ to provide any meaningful information about the flow structures at those frequencies.  Furthermore, figure~\ref{fig: htfAmpl} suggests
that the presence of higher harmonics is exclusively due to perturbations at the
fundamental frequency $\omega$ that scatter off the base flow to excite a
response at higher harmonics. It appears that nonlinear mechanisms
dominate the dynamics of this flow, and it is therefore necessary to perform a
linearization about a time-varying base flow in order to study the amplification
mechanisms.

\section{Conclusion}
In this paper we have considered small periodic perturbations about a
periodically time-varying base flow.  We have linearized the incompressible
Navier-Stokes equations about this time-varying base flow, and defined the
corresponding harmonic resolvent operator, a linear operator that describes the
evolution of these perturbations, including cross-frequency interactions.  In
particular, perturbations at frequency $\omega$ are coupled to perturbations at
frequency $\alpha$ through the base flow at frequency $\omega-\alpha$. If,
however, the dynamics are linearized about a steady base flow, as in the
standard resolvent framework, the coupling between structures at different
frequencies is lost.

We have shown that the right and left singular vectors of the harmonic resolvent
describe the dominant spatio-temporal amplification
mechanisms, for perturbations about the chosen base flow, and we showed how one can
quantify the cross-frequency interactions in the flow by analyzing the
block-singular values of the harmonic resolvent.  We illustrated the approach on
a three-dimensional toy
model, and then applied the
analysis to the flow over an airfoil at an angle of attack.  For this example,
the leading output mode (left singular vector) of the harmonic resolvent
operator accurately describes the flow
structures that develop in response to periodic forcing near the body. For this
example, linearizing about a periodic base flow is essential: if, by contrast,
one linearizes about a steady base flow as in the standard resolvent analysis,
inaccurate flow structures are obtained, and cross-frequency interactions cannot
be captured.

\section*{Acknowledgments}
The authors wish to thank Professor Luigi Martinelli (Princeton University) for the fruitful discussions related to approaching the solution of large, stiff linear systems such as the one that arises within harmonic transfer function analysis. We also wish to acknowledge Dr.\ Jonathan Halverson from Princeton High Performance Computing (HPC) for his assistance in optimizing the code for memory. All the large-scale computations were performed on Princeton University's Tiger cluster. This work was supported by the Air Force Office of Scientific Research, award FA9550-17-1-0084.

\bibliography{RowleyGroupReferences}
\bibliographystyle{jfm}

\end{document}